\documentclass[journal]{IEEEtran}   

\ifCLASSINFOpdf
\else
\fi

\newcommand{\dd}{\mathrm{d}}
\newcommand{\zerospace}{\setlength{\arraycolsep}{0pt}}
\usepackage{amsmath}
\usepackage{graphicx,psfrag} 
\usepackage{amsmath,amsthm,amsfonts,amssymb,bm} 

\usepackage[ruled]{algorithm2e}
\usepackage{algpseudocode}
\usepackage{xcolor}
\usepackage{dsfont}

\newcommand{\bs}{\boldsymbol}
\newtheorem{lemma}{Lemma}

\newtheorem{claim}{Claim}

\begin{document}

\title{
Estimation for High-Dimensional Multi-Layer Generalized Linear Model -- Part~I: The Exact MMSE Estimator
}

\author{
Haochuan Zhang,
Qiuyun~Zou*,
and
Hongwen~Yang
\thanks{
H. Zhang is with School of Automation, Guangdong University of Technology, Guangzhou 510006, China (haochuan.zhang@gdut.edu.cn).
}
\thanks{
Q. Zou was with School of Automation, Guangdong University of Technology, Guangzhou 510006, China, and is now with School of Information and Communication Engineering, Beijing University of Posts and Telecommunications, Beijing 100876, China (qiuyunzou@qq.com)
}
\thanks{
H. Yang is with School of Information and Communication Engineering, Beijing University of Posts and Telecommunications, Beijing 100876, China (yanghong@bupt.edu.cn).
}
\thanks{
*Corresponding author: Q.~Zou.
}

}

\maketitle

\begin{abstract}
This two-part work considers the minimum means square error (MMSE) estimation problem for a high dimensional multi-layer generalized linear model (ML-GLM), which resembles a feed-forward fully connected deep learning network in that each of its layer mixes up the random input with a known weighting matrix and activates the results via non-linear functions, except that the activation here is stochastic and following some random distribution. Part I of the work focuses on the exact MMSE estimator, whose implementation is long known infeasible. For this exact estimator, an asymptotic analysis on the performance is carried out using a new replica method that is refined from certain aspects. A decoupling principle is then established, suggesting that, in terms of joint input-and-estimate distribution, the original estimation problem of multiple-input multiple-output is indeed identical to a simple single-input single-output one subjected to additive white Gaussian noise (AWGN) only. The variance of the AWGN is further shown to be determined by some coupled equations, whose dependency on the weighting and activation is given explicitly and analytically. Comparing to existing results, this paper is the first to offer a decoupling principle for the ML-GLM estimation problem. To further address the implementation issue of an exact solution, Part II proposes an approximate estimator, ML-GAMP, whose per-iteration complexity is as low as GAMP, while its asymptotic MSE (if converged) is as optimal as the exact MMSE estimator.
\end{abstract}


\IEEEpeerreviewmaketitle

\section{Introduction}
This paper consider the problem of estimating high dimensional random inputs from their observations obtained from a multi-layer generalized linear model (ML-GLM) \cite{manoel2017multi}:
\begin{eqnarray}
\bs{y}
=
    \bs{f}_{L} \left( \bs{H}_{L}
    \cdots
    \bs{f}_{2} \left( \bs{H}_{2}
        \bs{f}_{1} (\bs{H}_{1} \bs{x}; \bs{\eta}_{1})
    ; \; \bs{\eta}_{2} \right)
    \cdots
    ; \; \bs{\eta}_{L} \right)
\end{eqnarray}
in which $\bs{x}$ the is high dimensional random input, $\bs{y}$ is the high dimensional observation, $\bs{H}_{\ell}$ is the weighting matrix in the $\ell$-th layer ($\ell=1,\ldots, L$) that linearly combines its input, and $\bs{f}_{\ell}(\bs{z}; \bs{\eta}_{\ell}) = \prod_{a=1}^{M} f_{\ell}(z_a; \eta_{\ell,a})$ is the activation function that maps the weighted result componentwisely.
The model, ML-GLM, resembles a feed-forward deep learning network of full connections in many aspects, except that the activation here is random. In particular, the activation here has a parameter $\bs{\eta}_{\ell}$ that follows some random distribution, and as a consequence, the entire activation process requires a transitional distribution to fully characterize its input-output relation. To see how this differs from a classical neural network, consider a case where the (deterministic) bias is replaced by some random values drawn from a Gaussian population. The activated result in that case is no longer deterministic due to the random bias, even though the activation in itself is deterministic.
The ML-GLM is a general model, embracing many well-known models as special cases. For instance, when $L=1$, it reduces to the generalized linear model (GLM) \cite{Rangan-arxiv10-GAMP, Rangan-ISIT11-GAMP}, a model described by $\bs{y}= \bs{f} (\bs{H} \bs{x}; \bs{\eta})$ and extensively adopted in low-resolution quantization studies \cite{He-JSTSP18-GEC_SR} where $\bs{y}=\text{ADC}(\bs{H} \bs{x}+\bs{\eta})$ with ADC($\cdot$) modeling the analog-to-digital conversion. As another instance, when the random activation is modeled by some additive white Gaussian noise (AWGN), the ML-GLM reduces to the celebrated standard linear model (SLM) \cite{donoho2009message}, where $\bs{y}= \bs{H} \bs{x}+ \bs{\eta}$ and its applications have a wide range of varieties, including wireless communications \cite{wen2014channel}, image processing \cite{metzler2017learned}, compressive sampling \cite{donoho2009message}, and many others.
As a generalization to the above models, the general ML-GLM is further able to model the inference problem arising in deep learning applications \cite{fletcher2018inference}.

For these models, the estimation problem is a classic yet still active topic, to which tremendous efforts had been dedicated during the past few decades. Among these is the minimum mean square error (MMSE) estimator, which is optimal in the MSE sense as its output $\hat{\bs{x}}$ could minimize $\mathbb{E} [\|\bs{x}-\hat{\bs{x}}\|^2]$.
The exact implementation of an MMSE estimator, however, is infeasible \cite{bishop2006pattern} (NP-hard) in high dimensional latent space, because of its requirements on the marginalization of a posterior distribution that contains many random variables or on the expectations over these distributions.
This issue was recognized as a facet of the curse of dimensionality, and as a remedy, people started to look at approximate solutions. Among those scaling well to high-dimensional applications, approximate message passing (AMP) \cite{donoho2009message} enjoyed a great popularity in scenarios with known and factorable priors. Originally designed for compressive recovery in the SLM setting, AMP was able to offer a Bayes-optimal estimation performance (it achieved the theoretical bound of a sparsity-undersampling tradeoff) but its implementation complexity is kept at a surprisingly low level (its message number scaled linearly with the variable number per iteration). Following AMP, a great number of approximate solutions had been proposed, and among these were three estimators pertaining to ML-GLM and thus are of particular interest here. The three are generalized AMP (GAMP) \cite{Rangan-arxiv10-GAMP}, multi-layer AMP (ML-AMP) \cite{manoel2017multi}, and multi-layer vector AMP (ML-VAMP) \cite{fletcher2018inference}. The first estimator, GAMP \cite{Rangan-arxiv10-GAMP}, extended AMP's scope (SLM) to allow non-linear activation (GLM); however, it considered only a single layer. As more recent advances, the latter two, ML-AMP \cite{manoel2017multi} and the ML-VAMP \cite{fletcher2018inference}, were able to handle the multi-layer case, but they also suffered from some limitations. In particular, the ML-VAMP \cite{fletcher2018inference} required a singular value decomposition (SVD) on each of the weighting matrices and thus inevitably comprised its computational efficiency, while the ML-AMP \cite{manoel2017multi}, although more efficient in computation (as no SVD needed), converged in a relatively slow speed (as one will see from our simulation section in Part II).

To fill in this gap, this two-part work proposes a new estimator, the ML-GAMP, whose convergence rate turns out to be faster than ML-AMP \cite{manoel2017multi} (by using messages that are more recently updated) and its computational burden is also lower than ML-VAMP \cite{fletcher2018inference} (since no SVD is required). In order to validate its optimality, we first analyze in Part I (i.e., this paper) the asymptotic performance of an exact MMSE estimator (despite of its implementation difficulty) by means of replica method \cite{parisi1979infinite, Mezard-Book87-SpinGlass_RM}, a powerful tool arising from statistical physics 30 years ago for attacking theoretical problems with sharp predictions. We derive the fixed point equations of the exact MMSE estimator, and compare them to the state evolution of the proposed estimator obtained in Part II. A perfect agreement is finally observed between the two, which suggests that the proposed estimator is able to attain asymptotically an MSE-optimal performance the same as the exact MMSE estimator. Since Part I of this work is dedicated to the (lengthy) replica analysis of an exact MMSE estimator, we leave all detail about the proposed ML-GAMP to Part II.
Below, we summarize two major findings of this Part I paper:
\begin{itemize}
  \item
A decoupling principle is established, revealing that, in terms of joint input-and-estimate distribution, the original estimation problem of multiple-input multiple-output (MIMO) nature is identical to a simple single-input single-output (SISO) system where only an effective additive white Gaussian noise (AWGN) is experienced. This decoupling principle, mostly inspired by the seminal work of Guo and Verdu \cite{GuoTIT2005RM}, substantially extends \cite{GuoTIT2005RM}'s result on SLM to allow for multi-layer cascading and non-linear activation. As $L=1$, it also degenerates smoothly to \cite{GuoTIT2005RM} in SLM and to \cite{schulke2016statistical} in GLM.

    \item
The noise variance in the SISO model above could be determined from the solution to a set of coupled equations, whose dependency on the weighting and the activation is given explicitly. Comparing to the most related work \cite{manoel2017multi}, important refinements are made to the classical method, and thus more details is revealed on the internal structure of the coupled equations, leading to an establishment of the decoupling principle.
\end{itemize}

\section{System Model and Problem Formulation} \label{sec:sysModel}
\begin{figure}[!t]
\centering
\includegraphics[width=0.5\textwidth]{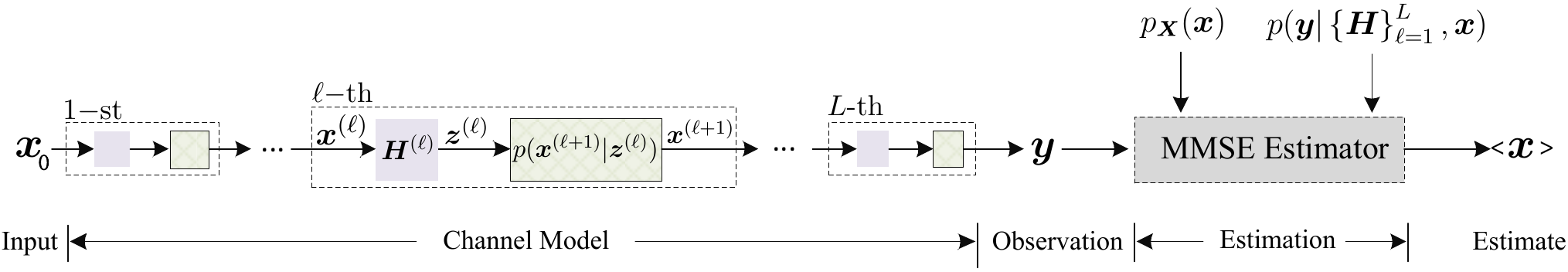}
\caption{System model of estimation in ML-GLM: random input $\to$ ML-GLM network $\to$ observation $\to$ MMSE estimator $\to$ output estimate.
}
\label{fig:mlglm}
\end{figure}
Fig.~\ref{fig:mlglm} is an illustration for the MMSE estimation in an ML-GLM setting. In the figure, $\bs{x}_0$ denotes the initial random input, whose distribution is factorable and known perfectly by the estimator, i.e.,
$
\bs{x}_0 \sim \mathcal{P}_{X}(\bs{x}_0)
=
    \prod_{i=1}^{N_1} \mathcal{P}_X (x_{0i})
,
$ 
$\bs{y}$ denotes the observation attained from the ML-GLM network  of $L$ layers, and $\langle \bs{x} \rangle$ is the output geneared by the MMSE estimator, in an manner either exact or approximate. Particularly, the $\ell$-th layer expands as ($1\leq \ell \leq L$)
\begin{equation}
\!\to\!\!
\bs{x}^{(\ell)}
\!\!\to\!\!
\boxed{ 
\bs{H}^{(\ell)}\bs{x}^{(\ell)}}
\!\!\to\!\!
\bs{z}^{(\ell)}
\!\!\to\!\!
\boxed {\mathcal{P}(\bs{x}^{(\ell+1)}|\bs{z}^{(\ell)})}
\!\!\to\! \bs{x}^{(\ell+1)}
\!\!\to\!\!
\label{A1}
\end{equation}
where $\bs{x}^{(\ell)}\in \mathbb{R}^{N_{\ell}}$ is its input, and $\bs{H}^{(\ell)}\in \mathbb{R}^{N_{\ell+1}\times N_{\ell}}$ is a deterministic weighting matrix that linearly mixes up the input to yield $\bs{z}\in \mathbb{R}^{N_{\ell+1}}$. This weighted result $\bs{z}$ is then activaed by a random mapping, whose transitional/conditional probability density function (p.d.f.) is also factorable:
$
\mathcal{P}(\bs{x}^{(\ell+1)}|\bs{z}^{(\ell)})
=
    \prod_{a=1}^{N_{\ell+1}}\mathcal{P}(x_a^{(\ell+1)}|z_a^{(\ell)})
.
$ 
The weighting matrix above is known perfectly to the estimator, and in each experiment, the elements of this matrix are drawn independently from the same Gaussian ensemble of zero mean and $1/N_{\ell+1}$ variance (to ensure a unit row norm).
To matain notational consistency, we also initialize: $\bs{x}^{(1)} := \bs{x}_0$, and $\bs{x}^{(L+1)} := \bs{y}$.
Since we consider exclusively the limiting performance of the MMSE estimators, the following assumptions are made throughout the paper: $N_{\ell}\to \infty$, while ${N_{\ell+1}}/{N_{\ell}} \to \alpha_{\ell}$, i.e., all weighting matrices are sufficiently large in size, but the ratios of their row numbers to culumn numbers are fixed and bounded%
\footnote{The ratio $\alpha_{\ell}$ could be either greater or smaller than $1$. For instance, in applications from wireless communications, it is usually the case $\alpha_{\ell} \geq 1$ for better signal recovery; while in compressed sensing applications, $\alpha_{\ell} \leq 1$ is desired to yield a better compression rate.
}%
.

The target of an exact MMSE estimator is to generate an estimate $\langle x_k\rangle$ for every input element $x_{0k}$ using ($k=1,\cdots,N_1$)
\begin{equation}
\langle x_k\rangle
=
\arg \min_{\hat{x}_k} \mathbb{E} \left[ \|\hat{x}_k-x_k\|^2 \right]
=\mathbb{E}\left[x_k \left|\bs{y},\{\bs{H}^{(\ell)}\} \right. \right]
\label{eq:MMSE_est}
\end{equation}
where the last expectation is taken over a marginal posterior
\begin{align}
\mathcal{P}(x_{0k}|\bs{y},\{\bs{H}^{(\ell)}\})=\int \mathcal{P}(\bs{x}_0|\bs{y},\{\bs{H}^{(\ell)}\})\dd \bs{x}_{0\backslash k}
,
\end{align}
whose integration is $(N_1-1)$-fold, $\bs{x}_{0\backslash k}$ equals $\bs{x}_0$ except its $k$-th element moved, and the joint p.d.f. $\mathcal{P}(\bs{x}_0|\bs{y},\{\bs{H}^{(\ell)}\})$ is:
\begin{align*}
\mathcal{P}(\bs{x}_0|\bs{y},\{\bs{H}^{(\ell)}\})
=
    \frac{
        \mathcal{P}_{X}(\bs{x}_0)\mathcal{P}(\bs{y}|\bs{x}_0,\{\bs{H}^{(\ell)}\})
    }
    {\int \mathcal{P}_{X}(\bs{x}_0)\mathcal{P}(\bs{y}|\bs{x}_0,\left\{\bs{H}^{(\ell)}\right\})\dd \bs{x}_0
    }
.
\end{align*}
For the above MMSE estimator, we note that its exact implementation requires the evaluation of a multi-fold integral as above. In high dimensional scenarios, this is apparently infeasible.
For performance analysis, we also note that, this two-part work adopts the average MSE defined as:
\begin{align}
    \mathrm{avgMSE}
     \triangleq
        \frac{1}{N_1} \sum\nolimits_{k=1}^{N_1} \mathbb{E} \left[ \| \langle x_k\rangle -x_k \|^2 \right]
     ,
\end{align}
i.e., an average of MSE realizations over all input $\bs{x}$, weighting $\{\bs{H}^{(\ell)}\}$, and activation $\{\mathcal{P}(x_a^{(\ell+1)}|z_a^{(\ell)})\}$ randomness.

Next, we start from a relatively simple case of $2$L-GLM to analyze the exact MMSE estimator's performance. Its result will be extended to more general cases in subsequent sections.

\section{Asymptotic Analysis for Two-Layer Case}\label{sec:2Layer}
\begin{figure}[!t]
\centering
\includegraphics[width=0.49\textwidth]{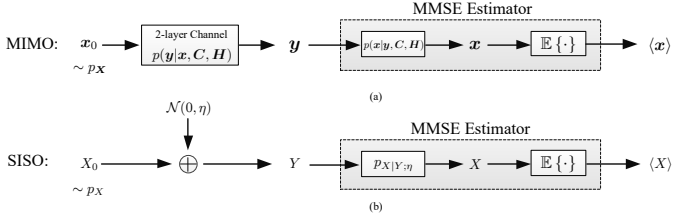}
\caption{The essential decoupling principle of MIMO to SISO (Claim \ref{claim:jointPDF})
}
\label{fig:RroChannel}
\end{figure}

To ease statement, we adopt a new set of notations in this two-layer section, hoping it to save us from the ocean of superscripts. To be specific, we re-denote 2L-GLM as below:
\begin{align*}
\footnotesize
\bs{x}_0\to
\underbrace{
\boxed{\bs{H}\bs{x}_0} \to \bs{u} \to \boxed{\mathcal{P}(\bs{s}|\bs{u})} \to \bs{s} \to \boxed{\bs{Cs}} \to \bs{v} \to \boxed{ \mathcal{P}(\bs{y}|\bs{v}) }
}
_{\mathcal{P}(\bs{y}|\bs{x}_0,\bs{C},\bs{H})}\to \bs{y}
\end{align*}
i.e., the general model (\ref{A1}) is particularized as:
$\bs{H}^{(1)} \leftarrow \bs{H}$,
$\bs{H}^{(2)} \leftarrow \bs{C}$,
$\bs{x}_0^{(2)} \leftarrow \bs{s}$,
$(N_1,N_2,N_3) \leftarrow (K,M,N)$,
$\alpha_1 \leftarrow \alpha$,
and
$\alpha_2 \leftarrow \beta$, while $K,M,N \to \infty$.
For simplicity, we define
\begin{equation}
\bs{u}
\triangleq
    \bs{H} \bs{x}_0
,
\quad
\bs{v}
\triangleq
    \bs{Cs}
,
\label{eq:def_v}
\end{equation}
with $\mathcal{P}(\bs{s}|\bs{u})$ and $\mathcal{P}(\bs{y}|\bs{v})$ denoting the new random activation in the two layers. The MMSE estimate of $\bs{x}_0$'s $k$-th element then becomes:
\begin{align}
\langle x_k\rangle =\mathbb{E}\left[ x_{0k}|\bs{y},\bs{C},\bs{H}\right]
,
\end{align}
where the expectation is taken over $\mathcal{P}(x_{0k}|\bs{y},\bs{H},\bs{C})$, i.e., the marginal of a joint posterior given by
$
\mathcal{P}(\bs{x}_0|\bs{y},\bs{C},\bs{H})=\frac{\mathcal{P}_{X}(\bs{x}_0)\mathcal{P}(\bs{y}|\bs{x}_0,\bs{C},\bs{H})}{\int \mathcal{P}_{X}(\bs{x}_0)\mathcal{P}(\bs{y}|\bs{x}_0,\bs{C},\bs{H}) \dd \bs{x}_0}
.
$
Under this new 2L-GLM notations, the system model is illustrated as Fig. \ref{fig:RroChannel}(a). Next, we present the main result from our replica analysis, while leaving its derivation details to the remainings subsections.

\subsection{Results for Exact MMSE Estimator in $2$L-GLM}
\label{subsec:2Layer_JM}

\begin{claim}[Joint distribution: 2-layer] \label{claim:jointPDF}
As illustrated in Fig. \ref{fig:RroChannel},
\begin{align}
\ (x_{0k},\langle x_{k}\rangle)
& \doteq
(X_0,\langle X \rangle)
,
\quad \forall k,
\end{align}
which means the exact MMSE estimation in a MIMO 2L-GLM:
\begin{align}
\bs{x}_0
\overset{\mathcal{P}(\bs{y}|\bs{x}_0,\bs{C},\bs{H})}{\longrightarrow}
\bs{y} 
\overset{\mathbb{E}_{\bs{x}_0 | \bs{y}}[\bs{x}_0] }{\longrightarrow}
\langle \bs{x} \rangle
\end{align}
is identical, in terms of joint input-and-estimate distribution, to that in a SISO setting with only an (effective) AWGN:
\begin{align}
X_0 \overset{+ W }{\longrightarrow}
Y 
\overset{\mathbb{E}_{X_0 | Y}[X_0] }{\longrightarrow}
\langle X \rangle
\end{align}
where
$X_0\sim \mathcal{P}_X(X_0)$ is a scalar input following the same distribution as an element $x_{0k}$ of the original vector $\bs{x}_0$, $Y=X_0+W$ is the scalar received signal only corrupted by an AWGN, $W\sim \mathcal{N}(W| 0,\eta)$, and $\langle X \rangle$ is the scalar MMSE estimate obtained via $\langle X \rangle \triangleq \mathbb{E}_{X_0|Y}[X_0]$ with
$
\mathcal{P}(X_0|Y)
=
\frac{\mathcal{P}_X(X_0)\mathcal{N}(Y|X_0,\eta)}{\int \mathcal{P}_X(X_0)\mathcal{N}(Y|X_0,\eta){\rm d}X_0}
.
$ 
The noise variance $\eta$ could be further determined from the solution to the coupled equations (\ref{eq:FP2Layer}) below, using the relation
$
\eta \triangleq {1}/{(2\tilde{d})}
.
$ 
Let $\sigma_X^2$ denote the variance of $\mathcal{P}_X(x)$,
$\mathcal{N}(\bs{x}|\bs{a},\bs{A})$ be a Gaussian density of mean $\bs{a}$ and covariance (matrix) $\bs{A}$,
${\rm D}\xi \triangleq \mathcal{N}(\xi|0,1){\rm d}\xi$ be a Gaussian measure,
and
$\mathcal{N}_{s|u}(a,A,b,B)\triangleq\mathcal{P}(s|u)\mathcal{N}(s|a,A)\mathcal{N}(u|b,B)$, the coupled equations then read
\begin{subequations}\label{eq:FP2Layer}
\begin{align}
c&=\sigma_X^2\\
e&=\int_s\int_u |s|^2 \mathcal{P}(s|u)\mathcal{N}\left(u|0,\frac{c}{\alpha}\right){\rm d}u{\rm d}s\\
h&=\int_y\int_{\xi} \frac{\left|\int_v v \mathcal{P}(y|v)\mathcal{N}\left(v|\sqrt{\frac{f}{\beta}}\xi,\frac{e-f}{\beta}\right){\rm d}v\right|^2}{\int_v \mathcal{P}(y|v)\mathcal{N}\left(v|\sqrt{\frac{f}{\beta}}\xi,\frac{e-f}{\beta}\right){\rm d}v}{\rm D}\xi{\rm d}y
\label{DD1}\\
\tilde{f}&=\frac{\beta(\beta h-f)}{2(e-f)^2}\\
q&=\int_{\zeta}\int_{\xi}\frac{\left|\int_s\int_uu\mathcal{N}_{s|u}(\zeta,\frac{1}{2\tilde{f}},\sqrt{\frac{d}{\alpha}}\xi,\frac{c-d}{\alpha}){\rm d}u{\rm d}s\right|}
{\int_s\int_u\mathcal{N}_{s|u}(\zeta,\frac{1}{2\tilde{f}},\sqrt{\frac{d}{\alpha}}\xi,\frac{c-d}{\alpha}){\rm d}u{\rm d}s}{\rm D}\xi{\rm d}\zeta\\
\tilde{d}&=\frac{\alpha (\alpha q-d)}{2(c-d)^2}
\label{eq:tilde_d}
\\
d&=\int_{\zeta} \frac{\left|\int_x x \mathcal{P}_X(x)\mathcal{N}(x|\zeta,\frac{1}{2\tilde{d}}){\rm d}x\right|^2}{\int_x \mathcal{P}_X(x)\mathcal{N}(x|\zeta,\frac{1}{2\tilde{d}}){\rm d}x}{\rm d}\zeta
\label{eq:d_fpEqu}
\\
f&=\int_{\zeta}\int_{\xi}\frac{\left|\int_s\int_us\mathcal{N}_{s|u}(\zeta,\frac{1}{2\tilde{f}},\sqrt{\frac{d}{\alpha}}\xi,\frac{c-d}{\alpha}){\rm d}u{\rm d}s\right|}
{\int_s\int_u\mathcal{N}_{s|u}(\zeta,\frac{1}{2\tilde{f}},\sqrt{\frac{d}{\alpha}}\xi,\frac{c-d}{\alpha}){\rm d}u{\rm d}s}{\rm D}\xi{\rm d}\zeta
\label{DD2}
\end{align}
\end{subequations}
\end{claim}
\noindent For this claim, we have four remarks below.

\underline{Remark 1}:
Claim \ref{claim:jointPDF} suggests that, from an end-to-end point of view, each input element of the original (self-interfering) MIMO system experiences in effect an SISO AWGN channel that appears to be interference-free. However, the presence of other input elements does have an impact on the estimation performance, and this impact is reflected in a rise of the noise level in the equivalent SISO model. In the literature, this effective noise level's inverse is called multi-user efficiency \cite{Verdu-Book98-MultiuserDetection} (in the context of wireless communications based on CDMA). The lower the efficiency is, the poorer its estimator performs. In case of SLM, this multiuser efficiency was shown by \cite{GuoTIT2005RM} to be upper bounded by the inverse of the actual (not effective) noise level, indicating that adding more users into a originally single-user system only deteriorates the overall estimation performance, which confirms a common sense that the multi-user interference do have some negative impacts. A quantitative description on this will be given later in  (\ref{III-A6}) of this paper, where $\eta=\sigma_w^2+\frac{1}{\alpha}\varepsilon (\eta)$ is the rising-up noise level, $\sigma_w^2$ is the level before rising, and $\frac{1}{\alpha}\varepsilon (\eta)$ is the additional loss caused by adding up the input number. For more discussions in the SLM case, we refer the interested readers to \cite{GuoTIT2005RM}, and for the more general ML-GLM case, an in-depth analysis is omitted here due to limited space and left to further studies.

\underline{Remark 2}:
The above claim was obtained from a replica analysis (with certain refinements), whose details will be given in subsequent subsections. Inside the wireless communication community, related pioneering work include \cite{tanaka2002statistical} by Tanaka, \cite{GuoTIT2005RM} by Guo and Verdu, and their collaboration \cite{Guo-Book09-Generic}, but considering a CDMA application. In a more recent line of works, the replica method was to apply to the analysis of compressive sensing \cite{Kabashima-09-CS} by Kabashima \emph{et al.}, MIMO \cite{Wen-TIT06-RM_corrMIMO} by Wen \emph{et al.}, and massive MIMO \cite{Wen-TSP15-JCD} by Wen \emph{et al.}. All these work, however, concentrated on a single-layer setup, and the more general (also more challenging) ML-GLM was not considered until a recent work \cite{manoel2017multi} by Manoel \emph{et al.}. Comparing to \cite{GuoTIT2005RM} and \cite{manoel2017multi}, Claim \ref{claim:jointPDF}, on one hand, substantially extends the decoupling principle established  by \cite{GuoTIT2005RM} in a single-layer linear setting (SLM) to the much more general setting (ML-GLM) of multiple layers and non-linearity. On the other hand, it provided more details about the inner structure of the fixed point equations, as compared to \cite{manoel2017multi}. The above extension from 1L-SLM to ML-GLM is no trivial work, because (as discussed later) a key step in \cite{GuoTIT2005RM} is to compute certain covariance matrices in an explicit way, see \cite[(93)-(123)]{GuoTIT2005RM}, however, the computation becomes almost impossible in the presence of a non-Gaussian and non-linear activation. A new formulation is essential needed to handle the situation. For ML-GLM, although the standard replica method was applied in \cite{manoel2017multi} to analyze the performance, establishing a MIMO-to-SISO decoupling principle from the results there is no easy task.  In particular, from the state evolution in \cite[Eqs. (11)-(12)]{manoel2017multi}, it is challenging to sort out an explicit and one-step-only expression for the dependency of the fixed point equations on the input distribution $\mathcal{P}_X(x)$ and the final output estimate $\langle X \rangle$. As an evidence, see \cite[Eq. (11)]{manoel2017multi}, where $X$ and $\langle X \rangle$ are related only implicitly, i.e., via the interim variables generated for the processing of the many-fold layers in the middle. In contrast, Claim \ref{claim:jointPDF} (see also (\ref{eq:<X>})) here provides a an explicit and one-step-only expression on the dependency, paving the way for the decoupling principle's establishment. One reason for such a difference may take deep root in the different handling of $\lim_{\tau \to 0} \frac{\partial}{\partial \tau} \max_P \min_Q f(\tau, P, Q)$.
Previously, traditional replica analysis interchanged the order of the limiting and the extreme-value operations so that the analytical tractability could be kept \cite{Mezard-Book87-SpinGlass_RM}:
\begin{eqnarray*}
\lim_{\tau \to 0} \frac{\partial}{\partial \tau} \max_P \min_Q f(\tau, P, Q)
=\lim_{\tau \to 0} \max_P \min_Q \frac{\partial}{\partial \tau} f(\tau, P, Q)
.
\end{eqnarray*}
However, such an interchange had seldom been justified, and counter examples around in the mathematical world, if the function $f$ is arbitrary. Noticing this, we follow a different procedure to handle the evaluation, which retains the analytical tractability but at the same time is also rigorous mathematically. It reads here
\begin{align}
\lim_{\tau \to 0} \frac{\partial}{\partial \tau} \max_P \min_Q f(\tau, P, Q)
&=
    \lim_{\tau \to 0} \frac{\partial}{\partial \tau} f(\tau, P^*, Q^*)
\label{eq:limPartialDeriv}
\end{align}
with $(P^*, Q^*)$ being a solution to the following equation set
\begin{equation}
\frac{\partial}{\partial P^*} f(0, P^* \!\!, Q^*)
= 0
,
\;
\frac{\partial}{\partial Q^*} f(0, P^* \!\!, Q^*)
= 0
,
\label{eq:partialQ}
\end{equation}
See (\ref{eq:parital_Tau}) and above for a more detailed discussion.
Starting from this evaluation, the derivation in our paper differs from the traditional approach, although we are still following the same replica analysis framework and making important symmetry assumptions (among others). Summing up, the above refinements made to the standard replica method plays a significant role in our analysis, not only because it improves the method's rigorousness, but also it open a new avenue to look more closely into the inner structure of the coupled equations, which finally leads to our finding of the decoupling principle.

\underline{Remark 3}:
The coupled equations in Claim \ref{claim:jointPDF} may have multiple solutions, which was recognized as \emph{phase coexistence} in the literature. In statistical physics, as the system's parameters change, the dominant solution of the system may switch from one coexisting solution to another (thus termed \emph{phase transition}), and the thermodynamically dominant solution is the one that gives a smallest free energy value \cite{Guo-Book09-Generic}. While in the wireless communications context, a solution carrying the most relevant operational meaning is the one that yields an optimal spectral efficiency \cite{GuoTIT2005RM}.

\underline{Remark 4}:
From the discussion around (\ref{eq:<X>}), two quantities from Claim \ref{claim:jointPDF}, $c$ and $d$,  have some interesting interpretation: $c=\mathbb{E}[X^2]$ equals the power of a single input element, and $d=\mathbb{E}[\langle X \rangle^2]$ equals the power of its corresponding estimate. A natural idea from this interpretation is that, to evaluate the average MSE of a system, one only needs to compute a simple subtraction $c-d$, i.e.,
\begin{eqnarray}
\mathrm{avgMSE}
 = c - d.
\label{eq:MMSE2Layer}
\end{eqnarray}
meaning that given $c$ and $d$, one could be saved from the trouble of time-consuming Monte Carlo simulations that mimic the entire process of data generation, ML-GLM processing, MMSE estimation, and even error counting. To prove (\ref{eq:MMSE2Layer}), we start from the average MSE's definition:
$
\mathrm{avgMSE}
\triangleq
    \mathbb{E}[(X-\langle X \rangle)^2]
\overset{(a)}{=}
    \mathbb{E}[X^2 - \langle X \rangle^2]
=
    \mathbb{E}[X^2] - \mathbb{E}[\langle X \rangle^2]
,
$ 
where (a) applies the orthogonality principle from MMSE estimators, which says that, $\left(\langle X \rangle-X \right)$, the error vector of the MSE-optimal estimator is orthogonal to any possible estimator, including $\langle X \rangle$ itself.
Given $c$ and $d$, eq. (\ref{eq:MMSE2Layer}) could give the average MSE without simulations; this is one side of the coin. On the other side, in case of $d$ is not available%
\footnote{
The value of $c$ is always known as it is the variance of an input $X$, whose density is given by $\mathcal{P}_X(x)$.
} %
, eq. (\ref{eq:MMSE2Layer}) could give $d = c - \mathrm{avgMSE}(\tilde{d})$, saving us from the two-fold integral of (\ref{eq:d_fpEqu}), where the dependency of the average MSE on a known quantity $\tilde{d})$, defined in (\ref{eq:tilde_d}), is explicitly given by $\mathrm{avgMSE}(\tilde{d})$. An analytical expression for the dependency is possible, e.g., if the prior $\mathcal{P}_X(x)$ takes a QPSK form, then the average MSE could be rewritten explicitly as \cite{guo2011estimation}: $\mathrm{avgMSE}(\tilde{d})=1-\int \tanh (2\tilde{d}+\sqrt{2\tilde{d}}z)\text{D}z$, recalling  $\eta=1/(2\tilde{d})$.
Given the average MSE, it is also possible to compute numerically other performance indices. Take the symbol error rate (SER) as an example, if the transmitted symbol $X$ is drawn from a QPSK constellation, then the conversion from MSE to SER could be expressed analytically as \cite[p. 269]{Proakis-book01-DigitalCommu}
$
\text{SER}=2Q(\sqrt{\eta})-[Q(\sqrt{\eta})]^2
,
$ 
where $Q(x)=\int_{x}^{+\infty}\text{D}z$ is the $Q$-function.
Other prior distributions like the square QAM constellations are also possible, with more details being found in  \cite[p. 279]{Proakis-book01-DigitalCommu}.

Next, we consider the proof for Claim \ref{claim:jointPDF}, but before proceeding further, we first notice that an easier way to prove the equivalence in distribution is to calculate the moments and demonstrate their equivalence in values. Since in most cases of our interest the moments are assumed uniformly bounded \cite[eq. (166)]{GuoTIT2005RM}, this moment-calculation approach is reliable as per Carleman's theorem \cite[p. 227]{feller2008introduction}, saying that a distribution is uniquely determined by all its moments. For this reason, we prove instead the following lemma.
\begin{lemma} [Joint moment: 2-layer]\label{lem:JointMoment}
it holds ($i, j = 0, 1, 2, \cdots$)
\begin{align}
\mathbb{E}_{x_{0k},\bs{y},\bs{C},\bs{H}}\left[ x_{0k}^i \langle x_k\rangle ^j\right]
&=
    \mathbb{E}_{X_0,Y}\left[X_{0}^i\langle X \rangle^j\right]
.
\label{eq:JointMoment}
\end{align}
\end{lemma}
\noindent For the proof of this lemma, we will dedicate two subsections in the remaining of this section. The first subsection serves as a skeleton, while the second offers more details on several key items of the first.

\subsection{Replica Analysis--Part 1: Introducing the Replicas} \label{subsec:Proof_Claim1}
Before reformulating the joint moment expression, we now briefly explain the concept of ``replicas'':
\\
1) The original system:
\begin{eqnarray}
\bs{x}_0 \to \bs{y} \to \bs{x} \to \langle \bs{x} \rangle
\end{eqnarray}
The standard MMSE processing from an input $\bs{x}_0$ to an output
$\langle \bs{x} \rangle$ is denoted as
where $\bs{x}$ is a random variable that generates the output via its first-order moment, i.e., $\langle \bs{x} \rangle = \mathrm{E}[\bs{x}| \bs{y}]$, with $\bs{x} | \bs{y} \doteq \bs{x}_0 | \bs{y}$,  and $\bs{x}_0 | \bs{y} \sim \mathcal{P}(\bs{x}_0 | \bs{y})$.
\\
2) The replicated system:
\begin{eqnarray}
\bs{x}_0 \to \bs{y}
&
\to
    \begin{cases}
        \bs{x}_1 \to \langle \bs{x} \rangle
    \\
         \bs{x}_2 \to \langle \bs{x} \rangle
    \\
        \cdots \to \cdots
    \end{cases}
&
\end{eqnarray}
This is done by adding to the original system some ``replicas'', which are indeed i.i.d. random vectors $\bs{x}_1, \bs{x}_2, \ldots, \bs{x}_{\tau} $ conditioned on $\bs{y}$ and the channel matrices $\bs{C}$ and $\bs{H}$. These replicas generate the same estimate as in the original system, i.e.,
$\langle \bs{x} \rangle = \mathrm{E}[\bs{x}_{a}| \bs{y}]$, with $\bs{x}_a | \bs{y} \doteq \bs{x} | \bs{y}$ ($a=1, 2, \cdots$).

First of all, introduce $\tau$ replicas and rewrite  (\ref{eq:JointMoment})'s l.h.s. as
\begin{align}
\text{l.h.s.} (\ref{eq:JointMoment})
&=\mathbb{E}_{x_{0k},\bs{y},\bs{C},\bs{H}}
\left[[x_{0k}^i\prod_{u=1}^j \langle x_{uk}\rangle \right]
\\
&=\mathbb{E}_{x_{0k},\bs{y},\bs{C},\bs{H}} \left[ x_{0k}^i  \mathbb{E}\left(\prod_{u=1}^jx_{uk}|\bs{y},\bs{C},\bs{H}\right)\right]
\label{III-A1}\\
&=
    \mathbb{E}_{\left\{x_{uk}\right\}_{u=0}^j,\bs{y},\bs{C},\bs{H}}
    \left[x_{0k}^i \prod_{u=1}^j{x_{uk}}\right]
\\
&=
    \frac{1}{K}
    \mathbb{E}_{\bs{x}_0, \{\bs{x}_a\}, \bs{y},\bs{C},\bs{H}}
    \left[ \sum_{k=1}^{K} x_{0k}^i \prod_{u=1}^j{x_{uk}}\right].
\label{III-A2}
\end{align}
where the last equality follows from a self-averaging property of the high-dimensional signals \cite{GuoTIT2005RM, Cakmak-ITW14-Self_avg_AMP}.
Next, we show that
{\zerospace\begin{eqnarray}
\text{r.h.s.}(\ref{III-A2})
& = &
\!
\lim\limits_{
    \substack{%
    \tau \to 0
    \\
    h \to 0}
    }
\!
\frac{\partial }{\partial h} \! \log \mathbb{E}_{\bs{x}_0,\bs{y},\bs{C},\bs{H}}
[\mathcal{Z}^{(\tau)}
\!
(\bs{y}, \!\bs{C}, \!\bs{H}, \!\bs{x}_0; \!h) ]
\label{AA3}
\\
\mathcal{Z}^{(\tau)}(\cdot)
&\triangleq &
    \mathbb{E}_{\{\bs{x}_a\}}
    \!\!
    \left[
    \exp ( \frac{h}{K} \!\! \sum_{k=1}^K x_{0k}^i \!\! \prod_{u=1}^{j} \!\!x_{uk} )
    \prod_{a=1}^{\tau}\mathcal{P}(\bs{y}|\bs{x}_a,\bs{C},\bs{H})
    \right]
\nonumber
\end{eqnarray}}
where  $\{\bs{x}_a\} \triangleq [\bs{x}_1,\cdots,\bs{x}_\tau]$.
The proof for (\ref{AA3}) starts from an expansion on its r.h.s.. Substituting $\mathcal{Z}^{(\tau)}(\cdot)$ back into the formula and evaluate the partial derivative at its limit yields
\begin{eqnarray}
\text{r.h.s.}(\ref{AA3})
&=&
    \frac{1}{K}
    \lim_{\tau \to 0}
    \mathbb{E}_{\bs{x}_0,\bs{y},\bs{C},\bs{H}, \{\bs{x}_a\}}
    \left[
            x_{0k}^i\prod\nolimits_{u=1}^{j}x_{uk}
            \cdot
\right.
\nonumber\\&&
\left. \quad
    \prod\nolimits_{a=1}^{\tau}\mathcal{P}(\bs{y}|\bs{x}_a,\bs{C},\bs{H})
   \right]
\label{AA4}
\end{eqnarray}
According to the Bayes law of total probability, we have
\begin{align*}
\mathcal{P}(\{\bs{x}_a\}|\bs{y},\bs{C},\bs{H})
& =
    \prod_{a=1}^{\tau}
    \frac{
        \mathcal{P}(\bs{x}_a)    \mathcal{P}(\bs{y}|\bs{x}_a,\bs{C},\bs{H})
    }{
        \mathcal{P}(\bs{y} | \bs{C},\bs{H})
    }
\end{align*}
Substituting it into (\ref{AA4}) further rewrites the r.h.s. as ($\tau \to 0$)
\begin{align*}
\text{r.h.s.}(\ref{AA4})
=&
    \frac{1}{K}
    \mathbb{E}_{\bs{x}_0,\bs{y},\bs{C},\bs{H}}
    [\int \dd \{\bs{x}_a\}  x_{0k}^i\prod_{u=1}^{j}x_{uk}
    \mathcal{P}(\{\bs{x}_a\}|\bs{y},\bs{C},\bs{H})]
\\
=&
    \frac{1}{K}
    \mathbb{E}_{\bs{x}_0, \{\bs{x}_a\}, \bs{y},\bs{C},\bs{H}}[x_{0k}^i\prod_{u=1}^{j}x_{uk} \cdot]
=
\text{r.h.s.}(\ref{III-A2})
\end{align*}
which completes the proof for (\ref{AA3}). So far, we have proved
\begin{equation}
\text{l.h.s.} (\ref{eq:JointMoment})
\!\!=\!\!
\lim\limits_{
    \substack{%
    \tau \to 0
    \\
    h \to 0}
    }
\!
\frac{\partial }{\partial h}
\!
\frac{1}{K}
\!\log
\!
\mathbb{E}_{\bs{x}_0,\bs{y},\bs{C},\bs{H}}
[\mathcal{Z}^{(\tau)} \!(\bs{y}, \!\bs{C}, \!\bs{H}, \!\bs{x}_0; \!h)]
\label{AA5}
\end{equation}

Then, based on (\ref{AA5}), we continue to evaluate
$\frac{1}{K}\log\mathbb{E}[\mathcal{Z}^{(\tau)} (\cdot)]$, using high-dimensional random matrix theories. The result then reads
\begin{align}
\frac{1}{K}
\log
&
\mathbb{E}_{\bs{x}_0,\bs{y},\bs{C},\bs{H}}
[\mathcal{Z}^{(\tau)} \!(\bs{y}, \!\bs{C}, \!\bs{H}, \!\bs{x}_0; \!h)]
\nonumber\\
=&
    \underset{\bs{Q}_S,\tilde{\bs{Q}}_S,\bs{Q}_X}{\text{Extr}}
    \left[
    \alpha \beta G^{(\tau)}(\bs{Q}_S)
    - \alpha \text{tr}(\bs{Q}_S\tilde{\bs{Q}}_S)
    \right.\nonumber\\ & \left. \quad\quad
    + \alpha G^{(\tau)}(\tilde{\bs{Q}}_S,\bs{Q}_X)-R^{(\tau)}(\bs{Q}_X;h)
    \right]
\label{AAA1}\\
\triangleq &
    \underset{\bs{Q}_S,\tilde{\bs{Q}}_S,\bs{Q}_X,\tilde{\bs{Q}}_X}{\text{Extr}} T(\bs{Q}_S,\tilde{\bs{Q}}_S,\bs{Q}_X,\tilde{\bs{Q}}_X; \tau, h)
\label{AAA}
\end{align}
where the proof for the result of (\ref{AAA1}) will be given immediately in next subsection (Step 1 and 2).
`\text{Extr}' denotes an extreme value operation.
$\bs{Q}_X$, $\tilde{\bs{Q}}_X$, $\bs{Q}_S$ and $\tilde{\bs{Q}}_S$ are all $(\tau+1)\times (\tau+1)$ matrices. $G^{(\tau)}(\bs{Q}_S)$ and  $G^{(\tau)}(\tilde{\bs{Q}}_S,\bs{Q}_X)$ are defined around (\ref{eq:def_G(Q_s)}).
$R^{(\tau)}(\bs{Q}_X;h)$ is the rate function of a density below:
\begin{equation}
\mathcal{P}(\bs{Q}_X;h)
\! \triangleq \!
    \mathbb{E}_{\bs{x}_0, \{\bs{x}_a\}}
    \!\!
    \left(
        \frac{h}{K} \sum_{k=1}^K x_{0k}^i\prod_{u=1}^{j}x_{uk}
        \!
        \prod\limits_{0\leq a\leq b}^{\tau} \delta_{a,b}
    \right)
\end{equation}
where $\delta_{a,b}\triangleq \delta\left(\prod\limits_{k=1}^K x_{ak}x_{bk} -K[\bs{Q}_X]_{ab}\right)$ with $[\bs{Q}_X]_{ab}$ being the $(a,b)$-th element of $\bs{Q}_X$.
This rate function could be given explicitly using the large deviation theory as \cite[B-VI]{GuoTIT2005RM}:
\begin{align}
R^{(\tau)}
(\bs{Q}_X;h)
&=
    \sup_{\tilde{\bs{Q}}_X}
    \left\{
    \text{tr}(\bs{Q}_X\tilde{\bs{Q}}_X)-\log M^{(\tau)}(\tilde{\bs{Q}}_X)
    -
\right.\nonumber\\ & \!\!\!\left.
    [\log M^{(\tau)}(\tilde{\bs{Q}}_X;h)-\log M^{(\tau)}(\tilde{\bs{Q}}_X;0) ]
    \right\}
,
\\
M^{(\tau)}(\tilde{\bs{Q}}_X;h)
& \triangleq
    \mathbb{E}_{\bs{x}}
    [ \exp (h x_0^i \prod_{u=1}^j x_u )
    \exp (\bs{x}^T\tilde{\bs{Q}}_X\bs{x} ) ]
,
\end{align}
where $\bs{x} \triangleq [x_0,x_1,\cdots,x_{\tau}]^T$, with $x_a \doteq x_{ak}$ for $a=0, \ldots, \tau$. So, we have seen
\begin{equation*}
\text{l.h.s.} (\ref{eq:JointMoment})
\!\!=\!\!
\lim\limits_{
    \substack{%
    \tau \to 0
    \\
    h \to 0}
    }
\!
\frac{\partial }{\partial h}
\underset{\bs{Q}_S,\tilde{\bs{Q}}_S,\bs{Q}_X,\tilde{\bs{Q}}_X}{\text{Extr}}
\!\!\!\!
T(\bs{Q}_S,\tilde{\bs{Q}}_S,\bs{Q}_X,\tilde{\bs{Q}}_X; \tau, h)
\end{equation*}

After that, we continue to simplify the r.h.s. of the last equality. But before evaluating the partial derivative $\frac{\partial }{\partial h}$ of $\frac{1}{K}\log\mathbb{E}[\mathcal{Z}^{(\tau)} (\cdot)]$, we differentiate $T(\cdot)$ first w.r.t. to its four matrix arguments and let all the derivatives to equal zero (as required by the extreme value operation). A set of (coupled) saddle point equations are then obtained, as given in (\ref{DA5}). To these equations, denoting their matrix-valued solutions are denoted by $\bs{Q}_S^*(\tau;h)$, $\tilde{\bs{Q}}_S^*(\tau;h)$, $\bs{Q}_X^*(\tau;h)$, and  $\tilde{\bs{Q}}_X^*(\tau;h)$, we find that, these solutions are indeed independent of $\tau$, and their values could be derived from $T(\bs{Q}_S,\tilde{\bs{Q}}_S,\bs{Q}_X,\tilde{\bs{Q}}_X; 0, h)$, as discussed earlier in (\ref{eq:limPartialDeriv})-(\ref{eq:partialQ}). Here we note that treating $\tau$ as an explicit argument of $T(\cdot)$ is essential and mathematical rigorous, which avoids a problematic exchange between $\lim_{\tau \to 0}$ and $\frac{\partial}{\partial h}$ in the classical replica method. Further assuming a replica symmetry structure (see Step 3 in next subsection), we parameterize the solution matrices and thus break down the saddle point equations of a matrix form to some scalar ones, which are then called fixed point equations and given in (\ref{eq:FP2Layer}). By that, we have:
$
\text{l.h.s.} (\ref{eq:JointMoment})
=
\lim_{h \to 0}
\frac{\partial }{\partial h}
T(\bs{Q}_S^*,\tilde{\bs{Q}}_S^*,\bs{Q}_X^*,\tilde{\bs{Q}}_X^*; 0, h)
.
$ 

Now, we are able to evaluate the partial derivative and its limit, which yields:
$
\lim_{h \to 0}
\frac{\partial }{\partial h}
T(\cdot)
 =
\lim_{h \to 0}
\frac{\partial }{\partial h}\log M^{(\tau)}(\tilde{\bs{Q}}_X^* ;h)
,
$
and that
\begin{eqnarray}
\text{l.h.s.} (\ref{eq:JointMoment})
&= &
    \frac{\mathbb{E}_{\bs{x}}[x^i\prod_{u=1}^j x_u\exp ( \bs{x}^T\tilde{\bs{Q}}_X^* \bs{x} )]}
    {\mathbb{E}_{\bs{x}}[\exp (\bs{x}^T\tilde{\bs{Q}}_X^* \bs{x} )]}
\label{AA6}
\end{eqnarray}
Based on the replica symmetric $\tilde{\bs{Q}}_X^*$, the r.h.s. of the above (\ref{AA6}) could be further interpreted as a joint moment of two scalar r.v.'s, i.e., $\mathbb{E}_{X_0, Y}\left\{X_0^i \langle X \rangle^j\right\}$, where $Y=X_0+W$ with $X_0 \sim \mathcal{P}_X(X_0)$, $W\sim \mathcal{N}(W|0,\eta)$, and $\langle X \rangle$ is the MMSE estimate of $X_0$, see Step 4.4 in next subsection for more detail. By that, an equivalent SISO AWGN model is established, completing the proof for:
$
\text{l.h.s.} (\ref{eq:JointMoment})
=
    \mathbb{E}_{X_0,Y}\left\{X_0^i \langle X \rangle^j\right\}
    .
$

\subsection{Replica Analysis--Part 2: Computing the Free Energy} \label{subsec:2Layer_FreeEnergy}
This subsection elaborates more details on the proof of some key steps skipped from last subsection to ease reading. These contents fit well into the framework of free energy computation for the replicated system, after noticing from (\ref{AA3}) that
$
\lim_{h \to 0} \mathcal{Z}^{(\tau)}(\bs{y}, \!\bs{C}, \!\bs{H}, \!\bs{x}_0; \!h)
=\mathcal{P}^{\tau}(\bs{y}|\bs{C},\bs{H})
.
$ 
Upon this, the free energy of the replicated system is defined as below
\begin{align}
\mathcal{F}
& \triangleq
    -\frac{1}{K}\mathbb{E}_{\bs{y},\bs{C},\bs{H}}\left\{\log \mathcal{P}(\bs{y}|\bs{C},\bs{H})\right\}
\label{eq:def_free_energy}
\end{align}
Recalling the fact that
$
\mathbb{E}(\log \Theta)
=
\lim\limits_{\tau\to 0}
\frac{\partial }{\partial \tau}\log \mathbb{E}(\Theta^{\tau})
$,
this free energy could be computed via
\begin{align}
\mathcal{F}
& =
    -\lim_{\tau\to 0} \frac{\partial }{\partial \tau}
    \mathcal{F}_{\tau}
\label{eq:replica}
\\
\mathcal{F}_{\tau}
& \triangleq
    \frac{1}{K}\log \mathbb{E}_{\bs{y},\bs{C},\bs{H}} \left\{\mathcal{P}^{\tau}(\bs{y}|\bs{C},\bs{H})\right\}.
\label{eq:def_Ftau}
\end{align}
Following the convention of replica method like \cite{Mezard-Book87-SpinGlass_RM}, we assume (\ref{eq:replica}) to be valid for all real-valued $\tau$ in the vicinity of $\tau=0$, and remains valid also for integers $\tau=1,2,\cdots$. The rigorous mathematical minds will immediately question the validity of this last assumption. In particular, the expression obtained for integer values may not be valid for real values in general. As a matter of fact   \cite{Guo-Book09-Generic},  the continuation of the expression to real values is not unique, e.g., $f(\tau) + \sin(\tau \pi)$ and $f(\tau)$ coincide at all integer $\tau$ for every function $f(\cdot)$. Nevertheless, as we shall see, the replica method simply takes the same expression derived for integer values of $\tau$, which is natural and straightforward in the problem at hand. The rigorous justification for the above assumption is still an open problem. Surprisingly, this continuation assumption, along with other assumptions sometimes very intricate on symmetries of solutions, leads to correct results in all non-trivial cases where the results are known through other rigorous methods, see \cite{bayati2011dynamics, Barbier-PNAS19-GAMP} for examples on the AMP and GAMP cases. In other cases, the replica method produces results that match well with numerical studies.

Before proceeding to the evaluation of $\mathcal{F}$, we reformulate first the partition function $\mathcal{P}(\bs{y}|\bs{C},\bs{H})$ using
\begin{align}
\mathcal{P}(\bs{y}|\bs{C},\bs{H})
&=\int_{\bs{s}} \mathcal{P}(\bs{y}|\bs{C},\bs{s})\int_{\bs{x}}\mathcal{P}(\bs{s}|\bs{H},\bs{x})\mathcal{P}(\bs{x})\dd \bs{x}\dd \bs{s}
\label{AA0} \\
&=\int_{\bs{s}} \left(\int_{\bs{u}}\mathcal{P}(\bs{y}|\bs{u})\delta(\bs{u}-\bs{Cs})\dd \bs{u}\right) \dd \bs{s} \times
\nonumber\\
&\quad \int_{\bs{x}}\left(\int_{\bs{v}}\mathcal{P}(\bs{s}|\bs{v})\delta(\bs{v}-\bs{Hx})\dd \bs{x}\right)\mathcal{P}(\bs{x})\dd \bs{x}
\label{AA1}
\end{align}
Comparing to the 1L-SLM considered in \cite{GuoTIT2005RM}, our challenges here in the 2L-GLM include: first, an extra layer of network exists which suffers from mixing interference (caused by the weighting) and non-linear activation; second, an activation that is non-Gaussian distributed. To handle these, our solution is:\\
1) Reformulate the network as a two-fold integral in (\ref{AA0}), so that a nested structure in the expression could be exploited to apply a ``divide-and-conquer'' strategy that starts backwardly from the last layer, treating previous ones as its prior.\\
2) Incorporate a Dirac-$\delta$ function into the non-linear activation process, see (\ref{AA1}), so that the non-AWGN random mapping could be separated from the linear deterministic weighting, which further paves way for the essential Gaussian approximation to the activation (non-linear and non-Gaussian).
\\
Following this line, we take 4 steps to compute the free energy.

\underline{\textbf{Step 1}}:
Gaussian approximation for
$ \mathbb{E}
[\mathcal{P}^{\tau}(\bs{y}|\bs{C},\bs{H})]$ of $\mathcal{F}_{\tau}$:
\begin{align}
&\mathbb{E}_{\bs{y},\bs{C},\bs{H}} [\mathcal{P}^{\tau}(\bs{y}|\bs{C},\bs{H})]
=
    \mathbb{E}_{\bs{C},\bs{H}}[\int_{\bs{y}} \mathcal{P}^{\tau+1}(\bs{y}|\bs{C},\bs{H})\dd \bs{y}]
\\
&=
    \mathbb{E}_{\bs{C},\bs{H}}[\int_{\bs{y}}\prod_{a=0}^{\tau}\int_{\bs{x}_a}\mathcal{P}(\bs{y}|\bs{x}_a,\bs{C},\bs{H})\mathcal{P}(\bs{x}_a)\dd \bs{x}_a\dd  \bs{y}]
\end{align}
Then, it holds
\begin{align}
&\mathbb{E}_{\bs{y},\bs{C},\bs{H}} \left\{\mathcal{P}^{\tau}(\bs{y}|\bs{C},\bs{H})\right\}
\nonumber\\
=&
    \mathbb{E}_{\bs{C},\bs{H}}
    \left\{\int_{\bs{y}}\prod_{a=0}^{\tau}\int_{\bs{s}_a}\left(\int_{\bs{v}_a}\mathcal{P}(\bs{y}|\bs{v}_a)\delta(\bs{v}_a-\bs{Cs}_a)\dd \bs{v}_a\right)\right.
\times \nonumber\\
&
    \left.\int_{\bs{x}_a}\left(\int_{\bs{u}_a}\mathcal{P}(\bs{s}_a|\bs{u}_a)\delta(\bs{u}_a-\bs{H}\bs{x}_a)\dd \bs{u}_a\right)\mathcal{P}(\bs{x}_a)\dd \bs{s}_a\dd \bs{y}\right\}
\nonumber\\
=&\mathbb{E}_{\bs{S}}\left\{\int_{\bs{y}}\int_{\bs{V}}\prod_{a=0}^{\tau}\mathcal{P}(\bs{y}|\bs{v}_a)\mathbb{E}_{\bs{C}}\left\{\delta(\bs{V}-\bs{CS})\right\}\dd \bs{V}\dd \bs{y}\right\}
\label{D4}
\end{align}
where the subscript $a$ refers to the replica number, e.g., $\bs{x}_a$ being $a$-th replica of $\bs{x}$, and the following definitions are used:
$\bs{X} \triangleq [\bs{x}_0,\cdots,\bs{x}_{\tau}]$,
$\bs{U} \triangleq [\bs{u}_0,\cdots,\bs{u}_{\tau}]$,
$\bs{S} \triangleq [\bs{s}_0,\cdots,\bs{s}_{\tau}]$, and
$\bs{V} \triangleq [\bs{v}_{0},\cdots,\bs{v}_{\tau}]$.
Moreover, the $(a,n)$-th element of $\bs{V}$ is denoted by
$
v_{an}
\triangleq
[\bs{Cs}_a]_n,
$
while the expectation in (\ref{D4}) is taken over
$
\mathcal{P}(\bs{S})
=
    \mathbb{E}_{\bs{X}} \left[
        \int_{\bs{U}} \mathcal{P}(\bs{S}|\bs{U})\mathbb{E}_{\bs{H}}\left\{\delta(\bs{U}-\bs{HX})\right\}\dd \bs{U}
    \right]
.
$ 

After that, this element becomes the sum of a large number of random variables. According to central limit theorem, $v_{an}$ could be approximated  by a Gaussian r.v. distributed as $\mathcal{N}(v| 0, \sum_{m=1}^M {s_{am}s_{bm}}/N)$  because
\begin{align}
\mathbb{E}_{\bs{C}}[v_{an}]
&=
    \mathbb{E}_{\bs{C}} [\sum_{m=1}^M c_{nm}s_{am} ]=0
\label{D5}
\\
\mathbb{E}_{\bs{C}}[v_{an}v_{bn}]
&=
    \mathbb{E}_{\bs{C}}
    [
        \sum_{m=1}^M \!\! c_{nm}s_{am} \!\! \sum_{m'=1}^M \!\! c_{nm'}s_{bm'}
    ]
\!\! = \!\!
    \sum_{m=1}^M \!\! \frac{s_{am}s_{bm}}{N}
\label{D6}
\end{align}
Letting $\bs{v}_n \triangleq [v_{0n}, v_{1n}, \ldots, v_{\tau n}]^T$ and applying Gaussian approximation, eq. (\ref{D4}) could be rewritten as
\begin{align}
\mathbb{E}
&_{\bs{y},\bs{C},\bs{H}}
\left\{\mathcal{P}^{\tau}(\bs{y},\bs{C},\bs{H})\right\}
=
\nonumber\\
&
    \mathbb{E}_{\bs{S}}\left[
    \int_{\bs{y}}\int_{\bs{V}}\prod_{a=0}^{\tau} \mathcal{P}(\bs{y}|\bs{v}_a)\prod_{n=1}^N\mathcal{N}(\bs{v}_n|\bs{0},\frac{\bs{S}^T\bs{S}}{N})\dd \bs{V}\dd \bs{y}
    \right]
\label{DD5}
\end{align}

\underline{\textbf{Step 2}}: Approximation to $\mathcal{F}_{\tau}$ as per large deviation theory: Letting
$
\bs{Q}_S \triangleq \frac{1}{M}\bs{S}^T\bs{S}
,
$ 
the density of $\bs{Q}_S$ could be given as
\begin{equation*}
\mathcal{P}(\bs{Q}_S)
=\mathbb{E}_{\bs{S}}\left[\prod\nolimits_{0\leq a\leq b}\delta\left(M[\bs{Q}_S]_{ab}-\sum\nolimits_{m=1}^M s_{am}s_{bm}\right)\right]
\end{equation*}
For this density function, there exists a correlation in $\bs{s}_a$ due to the linear weighting; fortunately, such a correlation will vanish as a consequence of the \emph{self-averaging effect} in large system limit. The self-average effect suggests that, in a large system, the random vector $\bs{u}_a$ will approximately be distributed as Gaussian with a zero mean and a covariance matrix of $\sigma_X^2\bs{H}\bs{H}^T$, whose limit is $ \frac{\sigma_X^2}{\alpha}\mathbf{I}$. On the other hand, the transitional distribution $\mathcal{P}(\bs{s}_a|\bs{u}_a)$ is an identical and element-wise random mapping, meaning that all elements in the vector $\bs{s}_a$ are i.i.d.. Together with the fact that $[\bs{Q}_S]_{ab}=\frac{1}{M}\sum_{m=1}^M s_{am}s_{bm}$, it is natural to have the large deviation theory come into play. This large deviation theory is a branch of statistical studies that offers many useful results for the limiting distribution of the sum of i.i.d. random variables. Particularly in our case, we find that the target p.d.f. $\mathcal{P}(\bs{Q}_S)$ could be represented via the rate function $R^{(\tau)}(\bs{Q}_S)$ \cite{wen2006asymptotic}
{\setlength{\arraycolsep}{0pt}
\begin{eqnarray*}
\mathcal{P}(\bs{Q}_S)
&=& \exp \left[-MR^{(\tau)}(\bs{Q}_S) \right]
\\
R^{(\tau)}(\bs{Q}_S)
& \triangleq& \sup_{\tilde{\bs{Q}}_S}
\left[
    \text{tr}(\bs{Q}_S\tilde{\bs{Q}}_S)
    -
        \log \mathbb{E}_{\bs{S}}\left[\exp\left({\text{tr}(\tilde{\bs{Q}}_S\bs{S}^T\bs{S})}\right)\right]
    /M
\right]
\end{eqnarray*}
}
Based on these results, we continue to simplify (\ref{DD5}) as
{\zerospace
\begin{eqnarray*}
&&\mathbb{E}_{\bs{y},\bs{C},\bs{H}} \left\{\mathcal{P}^{\tau}(\bs{y}|\bs{C},\bs{H})\right\}\\
&=&
    \int_{\bs{Q}_S} \!\! \int_{\bs{y}} \!\! \int_{\bs{V}}\!\! \prod_{a=0}^{\tau} \mathcal{P}(\bs{y}|\bs{v}_a)\prod_{n=1}^N\mathcal{N}(\bs{v}_n|\bs{0},\frac{1}{\beta}\bs{Q}_S)\mathcal{P}(\bs{Q}_S)\dd \bs{V}\dd \bs{y}\dd \bs{Q}_S\\
&=&
    \int_{\bs{Q}_S} \!\!\!\! \mathcal{P}(\bs{Q}_S)\dd \bs{Q}_S \prod_{n=1}^N
    \int_{y_n} \!\! \int_{\bs{v}_n} \!\!\prod_{a=0}^{\tau} \mathcal{P}(y|v_{an})\mathcal{N}(\bs{v}_n|\bs{0},\frac{1}{\beta}\bs{Q}_S)\dd \bs{v}_n\dd y_n
\\
&=&\int_{\bs{Q}_S}
\left(
    \int_{y} \!\! \int_{\bs{v}} \!\! \prod_{a=0}^{\tau} \mathcal{P}(y|v_a)\mathcal{N}(\bs{v}|\bs{0},\frac{1}{\beta}\bs{Q}_S)\dd \bs{v}\dd y
\right)^{N} \mathcal{P}(\bs{Q}_S)\dd \bs{Q}_S\\
&=&
    \int \dd \bs{Q}_S \,
    e^{
    N\log \left(
    \int_{y}\int_{\bs{v}}\prod_{a=0}^{\tau} \mathcal{P}(y|v_a)\mathcal{N}(\bs{v}|\bs{0},\frac{1}{\beta}\bs{Q}_S)\dd \bs{v}\dd y
    \right) -MR^{(\tau)}(\bs{Q}_S) }
\end{eqnarray*}
}
Upon this, we apply the Varadhan's theorem \cite[(22)]{touchette2011basic} to get the following \emph{Laplace approximation} or \emph{saddle-point approximation} to $\mathcal{F}_{\tau}$, whose definition was in (\ref{eq:def_Ftau})
\begin{align}
\mathcal{F}_{\tau}
&\overset{}{=}\sup_{\bs{Q}_S} \left\{\frac{N}{K}G^{(\tau)}(\bs{Q}_S)-\frac{M}{K}R^{(\tau)}(\bs{Q}_S)\right\}\\
&=\sup_{\bs{Q}_S}\inf_{\tilde{\bs{Q}}_S} \left\{\alpha \beta G^{(\tau)}(\bs{Q}_S)-\alpha \text{tr}(\bs{Q}_S\tilde{\bs{Q}}_S) +
\right.\nonumber\\
& \quad  \left. \frac{1}{K}\log \mathbb{E}_{\bs{S}}\left\{\exp\left({\text{tr}(\tilde{\bs{Q}}_S\bs{S}^T\bs{S})}\right)\right\}\right\}
\label{D7}
\\
G^{(\tau)}(\bs{Q}_S)
& \triangleq \log \int_{y}\int_{\bs{v}}\prod_{a=0}^{\tau} \mathcal{P}(y|v_a)\mathcal{N}(\bs{v}|\bs{0},\frac{1}{\beta}\bs{Q}_S)\dd \bs{v}
\dd y
\label{eq:def_G(Q_s)}
\end{align}

Similar to $v_{an}$ in (\ref{D5})-(\ref{D6}), the element  $u_{am}$ of the matrix $\bs{U}$ could be handled in an analogous way. In particular, we define
$
\bs{Q}_X \triangleq \frac{1}{K}\bs{X}^T\bs{X}
$ 
whose p.d.f. is then given by
\begin{equation}
\mathcal{P}(\bs{Q}_X)
=\mathbb{E}_{\bs{X}} [\prod\nolimits_{0\leq a\leq b}\delta (\sum\nolimits_{k=1}^{K}x_{ak}x_{bk}-K[\bs{Q}_X]_{ab} ) ]
\label{EE3}
\end{equation}
According to \cite[Theo. II.7.1]{Ellis-Book07-Varadhan_Theorem}, the probability measure of $\bs{Q}_X$ satisfies the Varadhand's theorem
with a rate function $R^{(\tau)}(\bs{Q}_X)$, and it holds
\begin{align}
\log
&
\mathbb{E}_{\bs{S}}
\left\{\exp\left({\text{tr}(\tilde{\bs{Q}}_S\bs{S}^T\bs{S})}\right)\right\}
=
    \log \int_{\bs{S}} \exp (\text{tr}(\tilde{\bs{Q}}_S\bs{S}^T\bs{S})) \dd \bs{S}
    \cdot
\nonumber\\&
    \int_{\bs{X}} \mathcal{P}(\bs{X})\dd \bs{X}
    \int_{\bs{U}} \mathcal{P}(\bs{S}|\bs{U})\mathbb{E}_{\bs{H}}\left\{\delta(\bs{U}-\bs{HX})\right\}\dd \bs{U}
\end{align}
Defining $\bs{u}_m \triangleq \left\{u_{am}\right\}_{a=0}^{\tau}$, it further breaks down as
\begin{align}
\log
&
\mathbb{E}_{\bs{S}}\left\{\exp\left(\cdot\right)\right\}
=
    \log \int_{\bs{Q}_X} \dd \bs{Q}_X \mathcal{P}(\bs{Q}_X) \prod_{m=1}^M  \int_{\bs{s}_m}\int_{\bs{u}_m}
\nonumber\\& \quad
    \exp \left(\bs{s}_m^T\tilde{\bs{Q}}_S\bs{s}_m\right)
    \mathcal{P}(\bs{s}_m|\bs{u}_m)\mathcal{N}(\bs{u}_m|\bs{0},\frac{\bs{Q}_X}{\alpha})
    \dd \bs{u}_m\dd \bs{s}_m
\nonumber\\
=&
    \log \!\!\int \!\! \dd \bs{Q}_X \mathcal{P}(\bs{Q}_X)
    [
        \int \!\! \dd \bs{s} \dd \bs{u}
        \exp({\bs{s}^T\tilde{\bs{Q}}_S\bs{s}})
        \mathcal{P}(\bs{s}|\bs{u})\mathcal{N}(\bs{u}|\bs{0},\frac{\bs{Q}_X}{\alpha})
    ]^M
\nonumber
\end{align}
Finally, by denoting $\bs{x} \triangleq \left[x_0,x_1,\cdots,x_{\tau}\right]^T$, we have
\begin{align}
&\log
\mathbb{E}_{\bs{S}}\left\{\exp\left(\cdot\right)\right\}
\!\!= \!\!
    K \! \sup_{\bs{Q}_X}\left[
        \alpha G^{(\tau)} \!( \tilde{\bs{Q}}_S,\bs{Q}_X)
        \!\! -\!\!
        R^{(\tau)} \! (\bs{Q}_X)
    \right]
\label{D8}
\\
&
G^{(\tau)}(\tilde{\bs{Q}}_S,\bs{Q}_X)
\triangleq
    \log \int \dd \bs{u}\dd \bs{s}
    \exp({\bs{s}^T\tilde{\bs{Q}}_S\bs{s}})\mathcal{P}(\bs{s}|\bs{u})\mathcal{N}(\bs{u}|\bs{0},\frac{\bs{Q}_X}{\alpha})
\nonumber\\
&
R^{(\tau)}(\bs{Q}_X)
\triangleq \!
    \sup_{\tilde{\bs{Q}}_X} \!\!
    \left[\text{tr}(\bs{Q}_X\tilde{\bs{Q}}_X)
    \!-\!
    \log \mathbb{E}_{\bs{x}}[\exp({\bs{x}^T\tilde{\bs{Q}}_X\bs{x}})]\right]
\end{align}
Combining (\ref{D7}) and (\ref{D8}) yields (`Extr' is the extreme value)
\begin{align}
\mathcal{F}_{\tau}
=&
    \!\!
    \underset{\bs{Q}_S,\tilde{\bs{Q}}_S,\bs{Q}_X,\tilde{\bs{Q}}_X}{\text{Extr}}
    \!\!
    \left[
    \alpha \beta G^{(\tau)}(\bs{Q}_S)
    -\alpha \text{tr}(\bs{Q}_S\tilde{\bs{Q}}_S)
    -
    \text{tr}(\tilde{\bs{Q}}_X\bs{Q}_X)
    +
    \right. \nonumber\\ & \left.
    \alpha G^{(\tau)}(\tilde{\bs{Q}}_S,\bs{Q}_X)
    +
    \mathbb{E}_{\bs{x}}\left\{\exp\left({\bs{x}^T\tilde{\bs{Q}}_X\bs{x}}\right)\right\}
    \right]
\label{eq:FreeEngery}
\\
\triangleq & \;\;
    \underset{\bs{Q}_S,\tilde{\bs{Q}}_S,\bs{Q}_X,\tilde{\bs{Q}}_X}{\text{Extr}}
    T(\bs{Q}_X,\tilde{\bs{Q}}_X,\bs{Q}_S,\tilde{\bs{Q}}_S)
\label{D9}
\\
= & \;\;
    T(\bs{Q}_X^*,\tilde{\bs{Q}}_X^*,\bs{Q}_S^*,\tilde{\bs{Q}}_S^*)
\label{D9-2}
\end{align}
where the last eqaulity uses $\cdot^*$ to differentiate an extreme point from a general  (matrix) argument.

\underline{\textbf{Step 3}}: Partial derivation for saddle points:
By taking partial derivatives of $T(\cdot)$ w.r.t. $\bs{Q}_X$, $\tilde{\bs{Q}}_X$, $\bs{Q}_S$, and $\tilde{\bs{Q}}_S$, we obtain the saddle point equations below:
\begin{subequations}\label{DA1}
\begin{align}
\tilde{\bs{Q}}_S&=\beta\frac{\partial G^{(\tau)}(\bs{Q}_S)}{\partial \bs{Q}_S}
\\
\bs{Q}_S&=\frac{\partial G^{(\tau)}(\tilde{\bs{Q}}_S,\bs{Q}_X)}{\partial \tilde{\bs{Q}}_S}\\
\tilde{\bs{Q}}_X&=\alpha \frac{\partial G^{(\tau)}(\tilde{\bs{Q}}_S,\bs{Q}_X)}{\partial \bs{Q}_X}\\
\bs{Q}_X&=\frac{\mathbb{E}_{\bs{X}}\left\{\bs{x}\bs{x}^T\exp\left({\bs{x}^T\tilde{\bs{Q}}_X\bs{x}}\right)\right\}}{\mathbb{E}_{\bs{x}}\left\{\exp\left({\bs{x}^T\tilde{\bs{Q}}_X\bs{x}}\right)\right\}}
\end{align}
\end{subequations}
To further simplify the matrix derivative, we find the following identity very useful (see the supporting materials for a proof):
\begin{align*}
\frac{\partial \mathcal{N}(\bs{x}|\bs{a},\bs{A})}{\partial \bs{A}}
=
\!\!
\frac{-1}{2}
[
    \bs{A}^{-1}
    \!\! - \!\!
    \bs{A}^{-1}(\bs{x}-\bs{a})(\bs{x}-\bs{a})^T \!\! \bs{A}^{-1}
]
\mathcal{N}(\bs{x}|\bs{a},\bs{A})
.
\end{align*}
By chain rule, this could rewrite the saddle point equations as
\begin{subequations}\label{DA5}
\begin{align}
\tilde{\bs{Q}}_S
&=-\frac{\beta}{2}(\bs{Q}_S^{-1}-\beta\bs{Q}_S^{-1}\mathbb{E}_{\bs{v}}[\bs{vv}^T]\bs{Q}_S^{-1})
\label{DA5:a}
\\
\bs{Q}_S
&=\mathbb{E}_{\bs{s}}[\bs{ss}^T]
\label{DA5:b}
\\
\tilde{\bs{Q}}_X
&=-\frac{\alpha}{2}\left(\bs{Q}_X^{-1}-\alpha\bs{Q}_X^{-1}\mathbb{E}_{\bs{u}}[\bs{uu}^T]\bs{Q}_X^{-1}\right)
\label{DA5:c}
\\
\bs{Q}_X
&=\frac{\mathbb{E}_{\bs{x}}\left\{\bs{x}\bs{x}^T\exp\left({\bs{x}^T\tilde{\bs{Q}}_X\bs{x}}\right)\right\}}{\mathbb{E}_{\bs{x}}\left\{\exp\left({\bs{x}^T\tilde{\bs{Q}}_X\bs{x}}\right)\right\}}
\label{DA5:d}
\end{align}
\end{subequations}
where the expectations are taken over these distributions
\begin{align*}
p_{\bs{V}}(\bs{v})
&=\frac{\int_y \prod_{a=0}^{\tau}\mathcal{P}(y|v^{(a)})\mathcal{N}(\bs{v}|0,\frac{1}{\beta}\bs{Q}_S)\dd y}
{\int_y \int_{\bs{v}}\prod_{a=0}^{\tau}\mathcal{P}(y|v^{(a)})\mathcal{N}(\bs{v}|0,\frac{1}{\beta}\bs{Q}_S)\dd \bs{v}\dd y}\\
p_{\bs{S}}(\bs{s})
&=\frac{\int_{\bs{u}}\exp(\bs{s}^T\tilde{\bs{Q}}_S\bs{s})\mathcal{P}(\bs{s}|\bs{u})\mathcal{N}(\bs{u}|\bs{0},\frac{\bs{Q}_X}{\alpha})\dd \bs{u}}
{\int_{\bs{s}}\int_{\bs{u}}\exp(\bs{s}^T\tilde{\bs{Q}}_S\bs{s})\mathcal{P}(\bs{s}|\bs{u})\mathcal{N}(\bs{u}|\bs{0},\frac{\bs{Q}_X}{\alpha})\dd \bs{u}\dd \bs{s}}\\
p_{\bs{U}}(\bs{u})
&=\frac{\int_{\bs{s}}\exp(\bs{s}^T\tilde{\bs{Q}}_S\bs{s})\mathcal{P}(\bs{s}|\bs{u})\mathcal{N}(\bs{u}|\bs{0},\frac{\bs{Q}_X}{\alpha})\dd \bs{s}}
{\int_{\bs{s}}\int_{\bs{u}}\exp(\bs{s}^T\tilde{\bs{Q}}_S\bs{s})\mathcal{P}(\bs{s}|\bs{u})\mathcal{N}(\bs{u}|\bs{0},\frac{\bs{Q}_X}{\alpha})\dd \bs{u}\dd \bs{s}}
\end{align*}
with $\mathcal{P}_{X}(\bs{x})$ being the prior density. On (\ref{DA5}), we note that it is in general very difficult to solve a four-matrix-argument solution $(\bs{Q}_X, \tilde{\bs{Q}}_X, \bs{Q}_S, \tilde{\bs{Q}}_S)$ out of the saddle point equations (\ref{DA5}), as there are too many arguments to solve. Although exceptions do exist, e.g., in case that all the prior and the transitional probabilities follow Gaussian distributions, the MMSE estimators there are usually simple to analyze and were thus extensively studied in the literature. For instance, in the all Gaussian case above, the exact MMSE particularize as the well known linear MMSE (LMMSE) estimator, whose asymptotic performance was well captured by the Tse-Hanly equations \cite{Tse1999Linear}. In this context, it is usually assumed that the solution will exhibit a certain pattern in the structure of each solution matrix, which is termed \emph{replica symmetry}. The replica symmetry considered here assumes that each matrix is a circular matrix consisting of two free parameters, thus reducing the number of individual equations from $4(\tau+1)^2$ to $4\times2$. It is worthy of noting that assuming replica symmetry, the free energy could be obtained analytically; however,
there is unfortunately no known general condition for the replica symmetry
to hold \cite{GuoTIT2005RM}%
\footnote{According to \cite{GuoTIT2005RM}, the validity of replica symmetry can be checked by calculating the Hessian of at the replica symmetric supremum \cite{nishimori2001statistical}. If the Hessian is positive definite, then the replica symmetric solution is stable against replica symmetry breaking, and it is the unique solution because of the convexity of the function. Under equal-power binary input and individually optimal detection, \cite{tanaka2002statistical} showed that if the system parameters satisfy certain condition, the replica-symmetric solution
is stable against replica symmetry breaking. In some other cases, replica symmetry can be broken \cite{Kabashima2003A}. Recently, Reeves and Pfister \cite{reeves2016replica} proved that the replica-symmetric prediction is exact for compressed sensing with Gaussian matrices.}%
. The replica-symmetric solution, assumed for analytical tractability in this paper, is consistent with numerical results in the simulation sections. In next step we provide more detail on the replica symmetric solutions.

\underline{\textbf{Step 4}}: Solutions under replica symmetry: Assuming replica symmetry, each solution matrix is parameterized by two free arguments, i.e., (for simplicity, we omit the superscript $^*$ despite of that fact that the variable is an extreme point, i.e., a solution to the saddle point equations)
\begin{subequations}\label{eq:replicaSymmetry}
\begin{align}
\bs{Q}_X&=(c-d)\mathbf{I}+d\bs{11}^T, \quad \tilde{\bs{Q}}_X=(\tilde{c}-\tilde{d})\mathbf{I}+\tilde{d}\bs{11}^T
\\
\bs{Q}_S&=(e-f)\mathbf{I}+f\bs{11}^T, \quad \tilde{\bs{Q}}_S=(\tilde{e}-\tilde{f})\mathbf{I}+\tilde{f}\bs{11}^T
\end{align}
\end{subequations}
with $(c,d,\tilde{c},\tilde{d},e,f,\tilde{e},\tilde{f})$ being the free parameters, and $\bs{11}^T$ denoting a all-one matrix of the size $(\tau+1)\times (\tau+1)$. For (\ref{DA5}), letting
$\bs{P}_1 \triangleq \mathbb{E}_{\bs{v}}[\bs{vv}^T]$ and $\bs{P}_2 \triangleq \mathbb{E}_{\bs{u}}[\bs{uu}^T]$, the two matrices  exhibit also some replica symmetry, so we have
\begin{align}
\bs{P}_1
&=(g-h)\mathbf{I}+h\bs{11}^T
\\
\bs{P}_2
&=(p-q)\mathbf{I}+q\bs{11}^T
\label{P2}
\end{align}
with $(g, h, p, q)$ being auxiliary parameters depending on $(c,d,\tilde{c},\tilde{d},e,f,\tilde{e},\tilde{f})$.
Using (\ref{D9}), one could rewrite the free energy as follows to emphasize explicitly its dependency on the eight parameters $(c,d,\tilde{c},\tilde{d},e,f,\tilde{e},\tilde{f})$ (as well as on the individual parameter $\tau$)
\begin{align}
\mathcal{F}=-\lim_{\tau\to 0}\frac{\partial }{\partial \tau} \mathcal{F}(\tau,c,d,\tilde{c},\tilde{d},e,f,\tilde{e},\tilde{f})
\label{DD3}
\end{align}
For this new expression, it is important to note that all the eight parameters of $\mathcal{F}(\tau,c,d,\tilde{c},\tilde{d},e,f,\tilde{e},\tilde{f})$ are actually functions of $\tau$, as one may recall from (\ref{eq:FreeEngery}) and (\ref{D7}) that the operations of $\mathrm{Extr}$ and $\sup \inf$ are carried out for a given $\tau$. In this regard, it is more precise to re-express the $\mathcal{F}_{\tau}$ term as   $\mathcal{F}(\tau,c_{\tau},d_{\tau},\tilde{c}_{\tau},\tilde{d}_{\tau},e_{\tau},f_{\tau},\tilde{e}_{\tau},\tilde{f}_{\tau})$.
Deriving the analytical results for all these eight $\tau$-dependent parameters are in general a challenging task, and thus a typical way adopted by statistical physicians for decades long is to exchange the partial derivative operation $\frac{\partial}{\partial \tau}$ outside the $\mathcal{F}_{\tau}$ term with the
$\mathrm{Extr}$ or $\sup \inf$ operation inside $\mathcal{F}_{\tau}$. Such an exchange is non-rigourous in general sense as counter examples abound in the mathematical world, though it had obtained great empirical successes during the years. In this paper, we apply a new approach to avoid such an exchange, and this approach in itself is rigourous in mathematical sense. Our approach is
\begin{align}
\mathcal{F}
&=
-\lim_{\tau\to 0}\frac{\partial }{\partial \tau} \mathcal{F}(\tau,c_{\tau},d_{\tau},\tilde{c}_{\tau},\tilde{d}_{\tau},e_{\tau},f_{\tau},\tilde{e}_{\tau},\tilde{f}_{\tau})
\\
&=
-\lim_{\tau\to 0}\frac{\partial }{\partial \tau} \mathcal{F}(\tau,c_{0},d_{0},\tilde{c}_{0},\tilde{d}_{0},e_{0},f_{0},\tilde{e}_{0},\tilde{f}_{0})
\label{eq:parital_Tau}
\end{align}
where the first equality uses an independent variable $\tau$ to highlight the explicit dependency, and the last equality is due to the fact given by (\ref{eq:limPartialDeriv})-(\ref{eq:partialQ}) that in overall effect, the free energy $\mathcal{F}$ depends only on its eight parameters evaluated at $\tau=0$. In other words, we don't have to solve out the parameters' expressions for arbitrary $\tau$ (and then perform a partial derivation followed by a limit); for the computation of free energy, we only need to solve them out at the origin point $\tau=0$. That is, we simply set $\tau=0$ in (\ref{DA5}) and then solve the coupled equations obtained. Our new approach is distinct from the (non-rigourous) conventional way in that we consider here jointly three operations, $\lim_{\tau\to 0}$, $\frac{\partial}{\partial \tau}$, and $\mathcal{F}_{\tau}$, but in classical way, it considers the latter two only, leaving it not too many choices except interchanging the two operations.

In the followings, we apply our new approach to solve/simplify the coupled equations. For notational simplicity, we abuse $(c,d,\tilde{c},\tilde{d},e,f,\tilde{e},\tilde{f})$ to denote
$(c_{0},d_{0},\tilde{c}_{0},\tilde{d}_{0},e_{0},f_{0},\tilde{e}_{0},\tilde{f}_{0})$, whenever their meanings are obvious from the context. The derivation is divided into four parts, detailed as below.

\noindent{\underline{\textbf{Step 4.1}}}: To solve $(\ref{DA5:a})$, we first evaluate  $g$ and $h$ as below
\begin{align}
g&=\frac{\int_y \int_{\bs{v}}(v_0)^2\prod_{a=0}^{\tau}\mathcal{P}(y|v_a)\mathcal{N}(\bs{v}|\bs{0},\frac{1}{\beta}\bs{Q}_S)\dd \bs{v}\dd y}
{\int_y \int_{\bs{v}}\prod_{a=0}^{\tau}\mathcal{P}(y|v_a)\mathcal{N}(\bs{v}|\bs{0},\frac{1}{\beta}\bs{Q}_S)\dd \bs{v}\dd y}\\
h&=\frac{\int_y \int_{\bs{v}}v_0v_1\prod_{a=0}^{\tau}\mathcal{P}(y|v_a)\mathcal{N}(\bs{v}|\bs{0},\frac{1}{\beta}\bs{Q}_S)\dd \bs{v}\dd y}
{\int_y \int_{\bs{v}}\prod_{a=0}^{\tau}\mathcal{P}(y|v_a)\mathcal{N}(\bs{v}|\bs{0},\frac{1}{\beta}\bs{Q}_S)\dd \bs{v}\dd y}
\end{align}
The key is to decouple $\bs{Q}_S$. Using the matrix inverse lemma, i.e., $(\bs{A}+\bs{BC})^{-1}=\bs{A}^{-1}-\bs{A}^{-1}\bs{B}(\mathbf{I}+\bs{CA}^{-1}\bs{B})^{-1}\bs{CA}^{-1}$, we have $\beta\bs{Q}_S^{-1}=\frac{\beta}{e-f}\mathbf{I}-\frac{f\beta}{(e-f)(e+f\tau)}\bs{11}^T$. Denote
$
A \triangleq \frac{\beta}{e-f}
,
$
$
B \triangleq \frac{f\beta}{(e-f)(e+f\tau)}
,
$ 
and evaluate
\begin{align}
&
\exp (-\frac{1}{2}\bs{v}^T\beta\bs{Q}_S^{-1}\bs{v})
=\exp\left[-\frac{A}{2}\sum_{a=0}^{\tau}v_a^2+\left(\sqrt{\frac{B}{2}}\sum_{a=0}^{\tau}v_a\right)^2\right]\nonumber\\
&\overset{(a)}{=}
\!\! \int \!\! \sqrt{\frac{\eta}{2\pi}}\exp \left[-\frac{A}{2}\sum_{a=0}^{\tau}v_a^2-\frac{\eta}{2}\xi^2 \!+\! \sqrt{\eta B}\sum_{a=0}^{\tau}v_a\xi\right] \!\! \dd \xi
\end{align}
where the last equality uses the \emph{Hubbard-Stratonovich transform} \cite{Hubbard1959-HS_Tranform}:
$
\exp \left(x^2 \right)=\sqrt{\frac{\eta}{2\pi}} \int_{\xi}\exp \left({-\frac{\eta}{2}\xi^2+\sqrt{2\eta}x\xi}\right)\dd \xi
$,
$\forall \eta>0$.
Now, we calculate $g$. Let  $C=(2\pi)^{-\frac{\tau+1}{2}}|\beta^{-1}\bs{Q}_S|^{-\frac{1}{2}}$, we have at $\tau\to 0$ (see supporting materials for a proof)
\begin{align}
&\int_y \int_{\bs{v}}\prod_{a=0}^{\tau}\mathcal{P}(y|v_a)\mathcal{N}(\bs{v}|\bs{0},\frac{\bs{Q}_S}{\beta})\dd \bs{v}\dd y
=C\sqrt{\frac{2\pi}{A-B}}
,
\label{DA6}
\\
&\int_y \int_{\bs{v}}v_0^2\prod_{a=0}^{\tau}\mathcal{P}(y|v_a)\mathcal{N}(\bs{v}|\bs{0},\frac{\bs{Q}_S}{\beta})\dd \bs{v}\dd y
=\frac{C}{A-B} \sqrt{\frac{2\pi}{A-B}}
,
\label{DA7}
\end{align}
Combining (\ref{DA6}) and (\ref{DA7}) yields
\begin{align}
g=\lim_{\tau\to 0}\frac{1}{A-B}=\frac{e}{\beta}
.
\end{align}
Defining $E=\frac{1}{e-f}$ and $F=\frac{f}{(e-f)(e+f\tau)}$, we have $\bs{Q}_S^{-1}=E\mathbf{I}-F\mathbf{11}^T$, and substituting $g=\frac{e}{\beta}$ into (\ref{DA5:a}), we further get
\begin{align}
\tilde{e}=\lim_{\tau\to 0} -\frac{\beta}{2}[(E-F)-\beta g(E-F)^2]=0
\label{DB2}
\end{align}

After that, we compute $h$ as $\tau \to 0$ (for more detail on the proof see the supporting materials of this paper)
\begin{align}
\int_y
&
\int_{\bs{v}}
v_0v_1\prod_{a=0}^{\tau}\mathcal{P}(y|v_a)\mathcal{N}(\bs{v}|\bs{0},\frac{1}{\beta}\bs{Q}_S)\dd \bs{v}\dd y
=\sqrt{\frac{2\pi}{A-B}} \times
\nonumber\\
&
\int_y \int_{\xi} \frac{\left[\int_v v \mathcal{P}(y|v)\mathcal{N}\left(v|\sqrt{\frac{B}{A(A-B)}}\xi,\frac{1}{A}\right)\dd v\right]^2}{\int_v  \mathcal{P}(y|v)\mathcal{N}\left(v|\sqrt{\frac{B}{A(A-B)}}\xi,\frac{1}{A}\right)\dd v}\text{D}\xi\dd y
\label{DB1}
\end{align}
which, together with (\ref{DA6}), yields
\begin{align}
h
&=\int_y\int_{\xi} \frac{\left[\int_v v \mathcal{P}(y|v)\mathcal{N}\left(v|\sqrt{\frac{f}{\beta}}\xi,\frac{e-f}{\beta}\right)\dd v\right]^2}{\int_v \mathcal{P}(y|v)\mathcal{N}\left(v|\sqrt{\frac{f}{\beta}}\xi,\frac{e-f}{\beta}\right)\dd v}\text{D}\xi\dd y
\end{align}

To evaluate $\tilde{f}$ of (\ref{DA5:a}), the following identity is useful as it indicates the existence of a replica-symmetry preserving property among the matrix product results: 
Given a $(\tau+1)\times (\tau+1)$ matrix $\bs{Q}=(a-b)\mathbf{I}_{\tau+1}+b\bs{11}^T$, it holds  \cite{shinzato2008perceptron}:
$
\bs{Q}=\bs{E}\left(
\begin{matrix}
a+\tau b &\bs{0}\\
\bs{0}  &(a-b)\mathbf{I}_{\tau}
\end{matrix}
\right)\bs{E}^T
,
$ 
where $\bs{E}=[\bs{e}_0,\cdots,\bs{e}_{\tau}]$ with $\bs{e}_0=[\frac{1}{\sqrt{\tau+1}},\cdots,\frac{1}{\sqrt{\tau+1}}]^T$ and the remaining being the $\tau$ orthogonal eigenvectors.
Given this, we rewrite (\ref{DA5:a}) as :
\begin{align}
\bs{G}_{\tilde{Q}_S}=-\frac{\beta}{2}(\bs{G}_{Q_S}^{-1}-\beta\bs{G}_{Q_S}^{-1}\bs{G}_{P_2}\bs{G}_{Q_S}^{-1})
\label{DB3}
\end{align}
where $\bs{G}_{Q_S}=
\left(
\begin{matrix}
e+\tau f &\bs{0}\\
\bs{0}  &(e-f)\mathbf{I}_{\tau}
\end{matrix}
\right)$ and
$\bs{G}_{P_2}=
\left(
\begin{matrix}
g+\tau h &\bs{0}\\
\bs{0}  &(g-h)\mathbf{I}_{\tau}
\end{matrix}
\right)$.
Combining (\ref{DB2})-(\ref{DB3}) yields ($\tau \to 0$)
\begin{align}
\tilde{f}=\frac{\beta(\beta h-f)}{2(e-f)^2}.
\end{align}

\noindent{\underline{\textbf{Step 4.2}}}: We next calculate (\ref{DA5:b}). By the Matrix Inversion Lemma, we see $\alpha \bs{Q}_X^{-1}=\frac{\alpha}{c-d}\mathbf{I}-
\frac{d\alpha}{(c-d)(c+d\tau)}\bs{11}^T$. Defining
 \begin{align}
 A'\triangleq\frac{\alpha}{c-d},\quad B'\triangleq\frac{d\alpha}{(c-d)(c+d\tau)}
 \end{align}
 and applying again the \textit{Hubbard-Stratonovich transform}, we decouple the tangled cross terms like $u_i u_j$ and $s_i s_j$ at the cost of an additional integral w.r.t. to a new auxiliary variable
\begin{align*}
&\exp\left(-\frac{1}{2}\bs{u}^T\alpha \bs{Q}_X^{-1}\bs{u}\right)
=
    \sqrt{\frac{\eta}{2\pi}} \int_{\xi} \dd \xi
\nonumber\\& \quad
\exp \left(-\frac{1}{2}A'\sum_{a=0}^{\tau}(u_a)^2-\frac{\eta}{2}\xi^2+\sqrt{\eta B'}\xi\sum_{a=0}^{\tau}u_a\right)
\\
&\exp \left(\bs{s}^T\tilde{\bs{Q}}_S\bs{s}\right)
=
    \sqrt{\frac{\gamma}{2\pi}}
    \int_{\zeta} \dd \zeta
\nonumber\\& \quad
\exp \left(-\tilde{f}\sum_{a=0}^{\tau}(s_a)^2-\frac{\gamma}{2}\zeta^2+\sqrt{2\gamma \tilde{f}}\zeta\sum_{a=0}^{\tau}s_a\right)
\end{align*}
With these decoupling results, $e$ now can be evaluated ($\tau \to 0$)
\begin{align}
&
\int
\exp(\bs{s}^T\tilde{\bs{Q}}_S\bs{s})\mathcal{P}(\bs{s}|\bs{u})\mathcal{N}(\bs{u}|\bs{0},\frac{\bs{Q}_X}{\alpha})\dd \bs{u}\dd \bs{s}
=
    C\sqrt{\frac{2\pi}{A'-B'}}
\label{DC1}
\\
&
\int
s_0^2\exp (\bs{s}^T\tilde{\bs{Q}}_S\bs{s})\mathcal{P}(\bs{s}|\bs{u})\mathcal{N}(\bs{u}|\bs{0},\frac{\bs{Q}_X}{\alpha})\dd \bs{u}\dd \bs{s}
= C' \sqrt{\frac{2\pi}{A'-B'}} \times
\nonumber\\
&
    \quad
    \int 
    s^2 \mathcal{P}(s|u)\mathcal{N}\left(u|\sqrt{\frac{B'}{A'(A'-B')}}\xi,\frac{1}{A'}\right)\dd u\dd s\text{D}\xi
\label{DC2}
\end{align}
where $C'=(2\pi)^{-\frac{\tau+1}{2}}|\alpha^{-1}\bs{Q}_X|^{-\frac{1}{2}}$, for more detail on the proof see the supporting materials. Combining (\ref{DC1})-(\ref{DC2}) yields
\begin{align}
e
=
\int_s\int_u |s|^2 \mathcal{P}(s|u)\mathcal{N}(u|0,\frac{c}{\alpha})\dd u\dd s
.
\end{align}

On the other hand, we have come to the simplification of $f$'s numerator (at $\tau \to 0$)
\begin{align}
&
\int_{\bs{s}}\int_{\bs{u}} s_0s_1\exp\left(\bs{s}^T\tilde{\bs{Q}}_S\bs{s}\right)\mathcal{P}(\bs{s}|\bs{u})\mathcal{N}(\bs{u}|\bs{0},\frac{\bs{Q}_X}{\alpha})\dd \bs{u}\dd \bs{s} = C' \times
\nonumber\\
&
\sqrt{\frac{2\pi}{A'-B'}}\int \!\!
\frac{\left|\int s\mathcal{N}_{s|u}(\zeta,\frac{1}{2\tilde{f}},\sqrt{\frac{d}{\alpha}}\xi,\frac{c-d}{\alpha})\dd u\dd s\right|^2}{\int \mathcal{N}_{s|u}(\zeta,\frac{1}{2\tilde{f}},\sqrt{\frac{d}{\alpha}}\xi,\frac{c-d}{\alpha})\dd u\dd s}\text{D}\xi\dd \zeta
\label{DC3}
\end{align}
which, together with (\ref{DC1}), further gives
\begin{align}
f=\int_{\zeta}\int_{\xi}\frac{\left|\int_{s}\int_{u} s\mathcal{N}_{s|u}(\zeta,\frac{1}{2\tilde{f}},\sqrt{\frac{d}{\alpha}}\xi,\frac{c-d}{\alpha})\dd u\dd s\right|^2}
{\int_{s}\int_{u} \mathcal{N}_{s|u}(\zeta,\frac{1}{2\tilde{f}},\sqrt{\frac{d}{\alpha}}\xi,\frac{c-d}{\alpha})\dd u\dd s}\text{D}\xi\dd \zeta
\label{DC5}
\end{align}
where $\mathcal{N}_{s|u}(a,A,b,B) \triangleq \mathcal{P}(s|u)\mathcal{N}(s|a,A)\mathcal{N}(u|b,B)$.

\noindent{\underline{\textbf{Step 4.3}}}: Before simplifying (\ref{DA5:c}), we still need $\bs{P}_2$, and we start from the numerator of $p$ as given in (\ref{P2}) (when $\tau \to 0$)
\begin{align}
&
    \int_{\bs{s}}\int_{\bs{u}}
    u_0^2\exp(\bs{s}^T\tilde{\bs{Q}}_S\bs{s})\mathcal{P}(\bs{s}|\bs{u})\mathcal{N}(\bs{u}|\bs{0},\frac{\bs{Q}_X}{\alpha})\dd \bs{u}\dd \bs{s}
\nonumber\\
=&
    C'
    \!\!
    \sqrt{\frac{2\pi}{A'-B'}}
    \!\!
    \int
    \!\!
    u^2
    \mathcal{N}_{s|u}
    \!\! \left( \!\!
    \zeta,\frac{1}{2\tilde{f}}, \sqrt{\frac{B'}{A'(A'-B')}}\xi,\frac{1}{A'}
    \!\!\right) \!\!
    \dd u\dd s\text{D}\xi\dd \zeta
\nonumber\\
=&
    C'\sqrt{\frac{2\pi}{A'-B'}}\frac{1}{A'-B'}
\label{DC4}
\end{align}
Combing (\ref{DC4}) and (\ref{DC1}), we get
\begin{align}
p=\lim_{\tau\to 0}\frac{1}{A'-B'}=\frac{c}{\alpha}.
\end{align}
For the simplification of $q$, we follow a procedure similar to that of $f$ in (\ref{DC3})-(\ref{DC5}), and the result is
\begin{align*}
q=\int_{\zeta}\int_{\xi}\frac{\left|\int_{s}\int_{u} u\mathcal{N}_{s|u}(\zeta,\frac{1}{2\tilde{f}},\sqrt{\frac{d}{\alpha}}\xi,\frac{c-d}{\alpha})\dd u\dd s\right|^2}
{\int_{s}\int_{u} \mathcal{N}_{s|u}(\zeta,\frac{1}{2\tilde{f}},\sqrt{\frac{d}{\alpha}}\xi,\frac{c-d}{\alpha})\dd u\dd s}\text{D}\xi\dd \zeta
\end{align*}
Defining $E=\frac{1}{c-d}$ and $F=\frac{d}{(c-d)(c+d\tau)}$, and substituting $p=\frac{c}{\alpha}$ into (\ref{DA5:c}), we get:
\begin{align}
\tilde{c}=-\frac{\alpha}{2}[(E-F)-\alpha p(E-F)^2]=0.
\end{align}
The simplification on $\tilde{d}$ is analogous to that of $\tilde{f}$ via the same matrix decomposition technique. Thus, we skip the detail and provide below its result:
\begin{align}
\tilde{d}=\frac{\alpha(\alpha q-d)}{2(c-d)^2}.
\end{align}

\noindent{\underline{\textbf{Step 4.4}}}: To establish the SISO equivalence, we recall that $\tilde{c}=0$, and apply the Hubbard-Stratonovich transform \cite{Hubbard1959-HS_Tranform, Stratonovich1957-HS_Tranform} to decouple a cross term arising in the simplification of $c$ and $d$, i.e.,
\begin{align}
&
\lim_{\tau\to 0}\mathbb{E}_{\bs{x}}\left[\exp\left(\bs{x}^T\tilde{\bs{Q}}_X\bs{x}\right)\right]
\nonumber\\
&=\int_{y}\sqrt{\frac{\eta}{2\pi}}
\int_x \exp \left[{-\frac{\eta}{2}(y-\sqrt{\frac{2\tilde{d}}{\eta}}x)^2}\right]\mathcal{P}_X(x)\dd x\dd y
\\
&=\int_x\int_{y}\mathcal{N}(y|x,\frac{1}{2\tilde{d}})\mathcal{P}_X(x)\dd x\dd y
=1
\label{DE3}
\end{align}
This further yields
\begin{align}
c&=\mathbb{E}_{\bs{x}}\left[x_0^2\exp \left(\bs{x}^T\tilde{\bs{Q}}_X\bs{x}\right)\right]
=
    \int X_0^2\mathcal{P}(X_0)\dd X_0
\label{DE1}\\
d&=\mathbb{E}_{\bs{x}}\left[x_0 x_1\exp \left(\bs{x}^T\tilde{\bs{Q}}_X\bs{x}\right)\right]=\int \langle X \rangle^2\mathcal{P}(Y)\dd Y
\label{DE2}
\end{align}
We interpret the distribution $\mathcal{N}(y|x,\frac{1}{2\tilde{d}})$ in the above equation a likelihood distribution of an observation $Y$ given the input $X_0$ in the context of a SISO system that reads
$
Y=X_0+W,
$
where $X_0 \sim \mathcal{P}_X(X_0)$, $W\sim \mathcal{N}(W | 0,\eta)$, $\eta = \frac{1}{2\tilde{d}}$, and $Y$ is the MMSE estimate of $X_0$, i.e.,
\begin{align}
\langle X \rangle=\int X_0\frac{\mathcal{N}(Y|X_0,\frac{1}{2\tilde{d}})\mathcal{P}_X(X_0)}{\int \mathcal{N}(Y|X_0,\frac{1}{2\tilde{d}})\mathcal{P}_X(X_0)\dd X_0}\dd X_0
\label{eq:<X>}
\end{align}
which establishes the SISO equivalence.

Given the solutions to the fixed point equations, i.e., $(c^*,d^*,\tilde{c}^*,\tilde{d}^*,e^*,f^*,\tilde{e}^*,\tilde{f}^*)$, we are now able to obtain the free energy $\mathcal{F}$ by substituting these solutions back to (\ref{eq:replicaSymmetry}) and later to (\ref{D9-2}), which completes the computation task.


\section{Asymptotic Analysis for $L$-Layer Case} \label{sec:L-Layer}

\subsection{ Results for Exact MMSE Estimator in ML-GLM}

\begin{claim}[Joint distribution: $L$-layer]
\label{claim:ML-jointPDF}
For the estimation in ML-GLM illustrated as Fig. \ref{fig:mlglm},
the exact MMSE estimation of a MIMO nature is identical, in the joint input-and-estimate distribution sense, to a simple SISO estimation  under an AWGN setting, i.e., ($k=1,\ldots, N_1$)
\begin{align}
(x_{0k},\langle x_{k}\rangle)
& \doteq
    (X_0,\langle X \rangle)
,
\end{align}
where $X_0$ and $\langle X \rangle$ are similarly defined as in Claim \ref{claim:jointPDF}, except the noise variance
$\eta ={1}/{(2\tilde{d})}$
is solved from Algorithm \ref{alg:Fixedpoint}.

%
\end{claim}

Claim \ref{claim:ML-jointPDF} indicates that, the existence of a SISO equivalence is not a sporadic phenomenon, but a universal truth that goes along with the multi-layer GLM. Such a ``decoupling property'' stands at the root of the replica method in statistical physics \cite{Barbier-PNAS19-GAMP}.
Owing to the generality of the multi-layer model, Claim \ref{claim:ML-jointPDF} embraces many existing results as its special cases, including:\\
1)  \underline{$L=2$, GLM:}
Claim \ref{claim:jointPDF} of this paper is a natural degeneration of Claim \ref{claim:ML-jointPDF} if one initializes $L$ as $2$ in Algorithm \ref{alg:Fixedpoint} and makes some trivial notation changes.
\\
2) \underline{ $L=1$, GLM \cite{schulke2016statistical}:}
In this case, the model degenerates to a (single-layer) generalized linear one, in which
Sch{\"u}lke \cite{schulke2016statistical} had shown the fixed point equations of an MMSE estimation result in the GLM could be written as follows
\begin{subequations}
\begin{align}
d&=\int_{\zeta} \frac{\left|\int_x xp_X(x)\mathcal{N}(x|\zeta,\frac{1}{2\tilde{d}}){\rm d}x\right|^2}{\int_x \mathcal{P}_X(x)\mathcal{N}(x|\zeta,\frac{1}{2\tilde{d}}){\rm d}x}{\rm d}\zeta
\\
q&=\int_{y}\int_{\xi}\frac{\left|\int_z z\mathcal{P}(y|z)\mathcal{N}(z|\sqrt{\frac{d}{\alpha}}\xi,\frac{\sigma_X^2-d}{\alpha}){\rm d}v\right|^2}{\int_z \mathcal{P}(y|z)\mathcal{N}(z|\sqrt{\frac{d}{\alpha}}\xi,\frac{\sigma_X^2-d}{\alpha}){\rm d }z}{\rm D}\xi {\rm d}y
\\
\tilde{d}&=\frac{\alpha (\alpha q-d)}{2(\sigma_X^2-d)^2}
\end{align}
\end{subequations}
which agrees perfectly%
\footnote{It is also worthy of noting that the above result is indeed a reproduction of \cite[(3.72)-(3.73)]{schulke2016statistical}, where one should pay special attention to the differences in our system setup, e.g., the weighting matrix is row normalized here while previously it was column normalized. In this context, the fastest way to verify this agreement is to consider a square weighting matrix.
} %
with Claim \ref{claim:ML-jointPDF} in case of $L=1$.
\\
3) \underline{ $L=1$, SLM \cite{GuoTIT2005RM, bayati2011dynamics}:}
The SLM is a further particularization of the GLM with $\mathcal{P}(y|v)=\mathcal{N}(y|v,\sigma_w^2)$. Substituting it back into the above GLM's fixed point equations, one gets a single-formula fixed point equation:
\begin{align}
\eta=\sigma_w^2+\frac{1}{\alpha}\varepsilon (\eta)
,
\label{III-A6}
\end{align}
where $\varepsilon (\eta)$ (as stated before) represents the average MSE of the AWGN channel, $Y=X+ W$, with $X \sim \mathcal{P}_X(x)$ and $W$ having a zero mean and a variance of $\eta$. This result was previously reported by \cite{GuoTIT2005RM} in the context of CDMA multiuser detection, and by  \cite{bayati2011dynamics} in the context of state evolution of AMP, another renowned statistical inference algorithm.


\begin{algorithm}[!t]
\caption{Fixed Point Equations of MMSE Estimator
}
\label{alg:Fixedpoint}
\setlength{\arraycolsep}{0pt}
\begin{eqnarray*}
\mathcal{P}^{(\ell)}(x|z) \triangleq \mathcal{P}_{x^{(\ell)}|z^{(\ell-1)}}(x|z)
,
\quad
\mathcal{N}^{(\ell)}_{x|z}(\cdot) \triangleq \mathcal{N}_{x^{(\ell)}|z^{(\ell-1)}}(\cdot)
\end{eqnarray*}
\\
\For{$\ell=1,\cdots,L$}
{
\begin{eqnarray*}
T_X^{(\ell)}
& = &
    \begin{cases}
    \ell=1:
    &
        \sigma_X^2
    \\
    \ell>1:
    &
        \int |x|^2\mathcal{P}^{(\ell)}(x|z)\mathcal{N}(z|0,\frac{T_X^{(\ell-1)}}{\alpha_{\ell-1}})\dd z\dd x
    \end{cases}
\end{eqnarray*}
}
\For{$\ell=L,\cdots,1$}
{
\begin{eqnarray*}
&&q^{(\ell)} =
\nonumber\\
&&\quad
    \begin{cases}
    \ell=L:
    \\
        \int \frac{\left|\int z \mathcal{P}^{(L)}(y|z)\mathcal{N} \left(z|\sqrt{\frac{d^{(L)}}{\alpha_{L}}} \xi,\frac{T_X^{(L)}-d^{(L)}}{\alpha_{L}}\right) \dd z\right|^2}{\int \mathcal{P}^{(L)}(y|z)\mathcal{N}\left(z|\sqrt{\frac{d^{(L)}}{\alpha_{L}}} \xi,\frac{T_X^{(L)}-d^{(L)}}{\alpha_{L}} \right)\dd z}\text{D}\xi\dd y
    \\
    \ell< L:
    \\
        \int\frac{\left|\int z\mathcal{N}^{(\ell)}_{x|z} \left(\zeta,\frac{1}{2\tilde{d}^{(\ell+1)}},\sqrt{\frac{d^{(\ell+1)}}{\alpha_{\ell}}} \xi,\frac{T_X^{(\ell+1)}-d^{(\ell+1)}}{\alpha_{\ell}}\right)
        \dd z\dd x\right|^2}{\int \mathcal{N}^{(\ell)}_{x|z}\left(\zeta,\frac{1}{2\tilde{d}^{(\ell+1)}},\sqrt{\frac{d^{(\ell+1)}}{\alpha_{\ell}}} \xi,\frac{T_X^{(\ell+1)}-d^{(\ell+1)}}{\alpha_{\ell}} \right)
        \dd z\dd x}\text{D}\xi\dd \zeta
    \end{cases}
\\
&&
\tilde{d}^{(\ell)}
=
    \frac{\alpha_{\ell}(\alpha_{\ell} q^{(\ell)}-d^{(\ell)})}{2(T_X^{(\ell)}-d^{(\ell)})^2}
\end{eqnarray*}
}

\For{$\ell=1,\cdots,L$}
{
\begin{eqnarray*}
&&d^{(\ell)} =
\nonumber\\
&&\quad
    \begin{cases}
    \ell=1:
    \\ \quad
        \int \frac{\left|\int x \mathcal{P}_X(x)\mathcal{N}\left(x|\zeta,\frac{1}{2\tilde{d}^{(1)}}\right)\dd x\right|^2}{\int \mathcal{P}_X(x)\mathcal{N}\left(x|\zeta,\frac{1}{2\tilde{d}^{(1)}}\right)\dd x}\dd \zeta
    \\
    \ell>1:
    \\
        \int \frac{\left|\int x \mathcal{N}^{(\ell)}_{x|z}\left(\zeta,\frac{1}{2\tilde{d}^{(\ell-1)}},\sqrt{\frac{d^{(\ell-1)}}{\alpha_{\ell-1}}}\xi,\frac{T_X^{(\ell-1)}-d^{(\ell-1)}}{\alpha_{\ell-1}}\right)
        \dd z\dd x\right|^2}
        {\int \mathcal{N}^{(\ell)}_{x|z} \left(\zeta,\frac{1}{2\tilde{d}^{(\ell-1)}},\sqrt{\frac{d^{(\ell-1)}}{\alpha_{\ell-1}}}\xi,\frac{T_X^{(\ell-1)}-d^{(\ell-1)}}{\alpha_{\ell-1}} \right)
        \dd z\dd x}\text{D}\xi\dd \zeta
    \end{cases}
\end{eqnarray*}

}
\end{algorithm}


\subsection{Sketch of Proof}
\label{MultiLayer_FreeEnergy}
Similar to Sec. \ref{subsec:2Layer_JM}, we will prove this moment identity:
\begin{align}
\mathbb{E}_{x_{0k},\bs{y},\{\bs{H}^{(\ell)}\}}
\left[x_{0k}^i\langle x_{k}\rangle ^j \right]
&=
    \mathbb{E}_{X_0,Y}
    \left[ X_0^i \langle X \rangle^j  \right]
\end{align}
First of all, we  notice that the discussions in Sec. \ref{subsec:Proof_Claim1} are indeed applicable to arbitrary $L$, so, for $L>2$, we only need to revisit its free energy computation.
To this end, we start all over again from the last layer and trace backward repeatedly until its very first, treating all previous layers as a prior to the current one. It begins with
\begin{equation*}
\mathcal{F}=-\frac{1}{K}\lim_{\tau\to 0}\frac{\partial }{\partial \tau} \log \mathbb{E}_{\bs{y},\{\bs{H}^{(\ell)} \}} \left[\mathcal{Z}^{\tau}(\bs{y},\{\bs{H}^{(\ell)}\})\right]
\end{equation*}
where $\mathcal{Z}(\bs{y},\{\bs{H}^{(\ell)}\})=\mathcal{P}(\bs{y}|\{\bs{H}^{(\ell)}\})$ is the partition function in the ML-GLM setting, and the expectation further expands
\begin{align}
&
\mathbb{E}_{\bs{y},\{\bs{H}^{(\ell)}\}} \left\{\mathcal{Z}^{\tau}(\bs{y},\{\bs{H}^{(\ell)}\})\right\}
=
    \mathbb{E}_{\bs{X}^{(L)}} \left\{\int \dd \bs{Z}^{(L)}\dd \bs{y}
\right.\nonumber\\ & \quad\left.
\prod_{a=0}^{\tau} \mathcal{P}(\bs{y}|\bs{z}_a^{(L)}) \times
\mathbb{E}_{\bs{Z}^{(L)}}\left[\delta(\bs{Z}^{(L)}-\bs{H}^{(L)}\bs{X}^{(L)})\right]\right\}
\label{IV-1}
\end{align}
where $\bs{z}_a^{(L)}$ denotes the $a$-th replica in the $L$-th layer (i.e., the last). We also have $\mathcal{P}(\bs{X}^{(1)})=\mathcal{P}(\bs{X})$, and for $\ell=L,\cdots,2$,
\begin{align}
\mathcal{P}(\bs{X}^{(\ell)})=
&
    \mathbb{E}_{\bs{X}^{(\ell-1)}} \left\{\int_{\bs{Z}^{(\ell-1)}} \dd \bs{Z}^{(\ell-1)}\mathcal{P}(\bs{X}^{(\ell)}|\bs{Z}^{(\ell-1)})
\times\right.\nonumber\\ &\left.
    \mathbb{E}_{\bs{H}^{(\ell-1)}}\left[\delta(\bs{Z}^{(\ell-1)}-\bs{H}^{(\ell-1)}\bs{X}^{(\ell-1)})\right]\right\}
\label{IV-2}
\end{align}

Next, we handle (\ref{IV-1}) and (\ref{IV-2}) in an analogous way to (\ref{D5})-(\ref{D7}) and (\ref{EE3})-(\ref{D8}), respectively. Then, the following saddle-point equations could be obtained ($\ell=L,\cdots, 1$)
\begin{subequations}
\begin{align}
\tilde{\bs{Q}}_X^{(\ell)}&=\frac{-\alpha_{\ell}}{2}\left[[\bs{Q}_X^{(\ell)}]^{-1}
\label{IV-3}
-\alpha_{\ell}[\bs{Q}_X^{(\ell)}]^{-1}\mathbb{E}\left(\bs{z}^{(\ell)}[\bs{z}^{(\ell)}]^T\right)[\bs{Q}_X^{(\ell)}]^{-1}\right]\\
\bs{Q}_X^{(\ell)}&=\mathbb{E}_{\bs{x}^{(\ell)}}\left(\bs{x}^{(\ell)}[\bs{x}^{(\ell)}]^T\right)
\label{IV-4}
\end{align}
\end{subequations}
with the expectations being taken over
\begin{align}
\mathcal{P}
&
_{\bs{Z}^{(\ell)}}
(\bs{z}^{(\ell)})
=
\nonumber\\
&
\frac{\int \exp \left(\bs{x}\tilde{\bs{Q}}_X^{(\ell+1)}\bs{x}\right)\mathcal{P}(\bs{x}|\bs{z}^{(\ell)})\mathcal{N}(\bs{z}^{(\ell)}|\bs{0},\frac{1}{\alpha_{\ell}}\bs{Q}_X^{(\ell)})\dd \bs{x}}
{\int \exp \left(\bs{x}\tilde{\bs{Q}}_X^{(\ell+1)}\bs{x}\right)\mathcal{P}(\bs{x}|\bs{z})\mathcal{N}(\bs{z}|\bs{0},\frac{1}{\alpha_{\ell}}\bs{Q}_X^{(\ell)})\dd \bs{z}\dd \bs{x}}
\\
\mathcal{P}
&
_{\bs{X}^{(\ell)}}
(\bs{x}^{(\ell)})
=
\nonumber\\
&
\frac{\int \exp \left(\bs{x}^{(\ell)}\tilde{\bs{Q}}_X^{(\ell)}\bs{x}^{(\ell)}\right)\mathcal{P}(\bs{x}^{(\ell)}|\bs{z})\mathcal{N}(\bs{z}|\bs{0},\frac{\bs{Q}_X^{(\ell-1)}}{\alpha_{\ell-1}})\dd \bs{z}}
{\int \exp \left(\bs{x}\tilde{\bs{Q}}_X^{(\ell)}\bs{x}\right)\mathcal{P}(\bs{x}|\bs{z})\mathcal{N}(\bs{z}|\bs{0},\frac{\bs{Q}_X^{(\ell-1)}}{\alpha_{\ell-1}})\dd \bs{z}\dd \bs{x}
}
\end{align}
where
$\tilde{\bs{Q}}_X^{(L+1)}=\bs{O}$,
and $\bs{Q}_X^{(0)}
=\frac{\mathbb{E}_{\bs{x}}\left\{\bs{x}\bs{x}^T\exp \left(\bs{x}^T\tilde{\bs{Q}}_X\bs{x}\right)\right\}}
{\mathbb{E}_{\bs{x}}\left\{\exp \left(\bs{x}^T\tilde{\bs{Q}}_X\bs{x}\right)\right\}}
$.

After that, assuming the solutions to the saddle-point equations exhibits the so-called replica symmetry, we compute the following items one by one: first, (\ref{IV-3}) using Step 4.1 as $\ell=L$ and using Step 4.3 as $\ell<L$; then, (\ref{IV-4}) using Step 4.2 as $\ell>1$ and Step 4.4 as $\ell=1$.

Finally, we get the fixed-point equations of Algo. \ref{alg:Fixedpoint} after some algebraic manipulations.

\subsection{Extension to Complex-Valued Settings} \label{sec:complex}

\begin{figure}[!t]
\centering
\includegraphics[width=.5\textwidth]{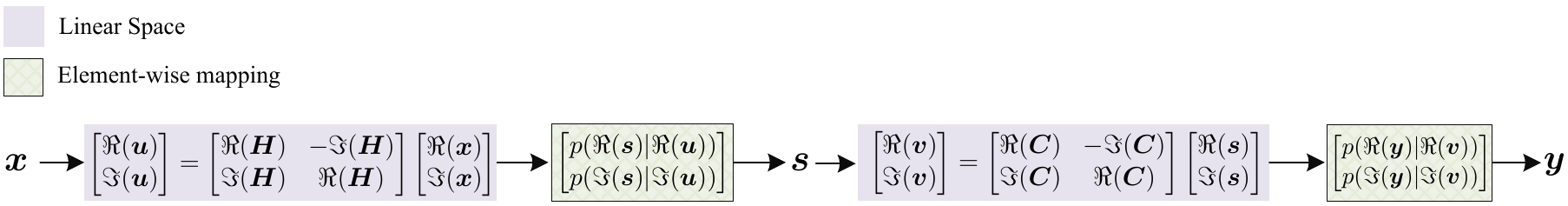}
\caption{The augmented matrix representation for complex-valued case.
}
\label{fig:Complex_2layerGLM}
\end{figure}

Until now the discussion has been based on a real-valued setting of the ML-GLM system, in which both the inputs and the transform matrix take real values. In practice, particularly in wireless communication systems like 5G,  spectral efficiency is a major concern, and the transmission is usually designed to be complex. In this section, we consider the extension of previous analysis to the complex settings. We follow \cite[Sec. V]{GuoTIT2005RM} to divide our discussion into 4 different cases: (a) real-input, real-transform; (b) complex-input, real-transform; (c) real-input, complex-transform;  (d) complex-input, complex-transform.
Since case (a) has already been studied in previous sections, we start from the second one.

In case (b), the inputs take complex values but the transform matrix is still real-valued. In this case, the system can be regarded as two uses of the real-valued transformations, where the inputs and the two transformations may be dependent. Since independent inputs maximize the channel capacity, there is little reason to transmit dependent signals in the two sub-systems. Thus, the analysis of the real-valued transform matrices in previous sections also applies to the case of independent in-phase and quadrature components, while the only change is that the spectral efficiency is the sum of that of the two sub-systems \cite[Sec. V]{GuoTIT2005RM}.

In case (c), the inputs take real values, while the transform matrix is complex. Comparing the complex-valued transformation to the real-valued one, it is easy to see that the complex-valued setting is equivalent to transmitting the same real-valued input twice over the two component real-valued channels. In other words, it is equivalent to having a real-valued channel with the load halved but input power doubled, in which our previous analysis is still applicable  \cite[Sec. V]{GuoTIT2005RM}.

In case (d), both the input and the transform matrix are complex-valued. The system model in this case could still be rewritten into an all real-valued one using the relationship between real and complex representations. We depict this new model in Fig. \ref{fig:Complex_2layerGLM}, where complex signals are reexpressed as real vectors/matrices and then mapped via the equivalent real-valued transformation. It appears that the previous analysis is not applicable to this new model as the transformation matrices here are not i.i.d. in their elements. However, as pointed out by \cite[Sec. V]{GuoTIT2005RM}, a closer look into the case, one would find that it is still possible to reuse the previous analysis after certain modifications. A key point here is that the variables $\bs{u}$ and $\bs{v}$ as defined around (\ref{eq:def_v}) have asymptotically independent real and imaginary components. Such an independency allows $G^{(\tau)}(\tilde{\bs{Q}}_S,\bs{Q}_X)$ and $G^{(\tau)}(\bs{Q}_S)$ as defined around (\ref{eq:def_G(Q_s)}) to be evaluated in analogy to the previous analysis. It turn outs that these two terms are doubled, comparing to the previous analysis. We also notice that if we assume the same signal power for both the real and the complex settings, then the real and the imaginary components in the complex case will both see a one-half power reduction%
\footnote{This is different from \cite[Sec. V]{GuoTIT2005RM}, where the signal power in the complex setting was doubled, and the situation for $G$ there was similar.}%
, which later balances out the doubling in $G^{(\tau)}(\tilde{\bs{Q}}_S,\bs{Q}_X)$ and $G^{(\tau)}(\bs{Q}_S)$ and leads to the final conclusion:
\emph{Given the same signal power, Claim \ref{claim:ML-jointPDF} is applicable to both the real and the complex ML-GLM's. }

\section{Conclusions} \label{sec:concl}

In this two-part work, we considered the problem of MMSE estimation for a high dimensional random input under the ML-GLM. As Part I of the two, this paper analyzed the asymptotic behavior of an exact MMSE estimator through the use of replica method. The replica analysis revealed that:
1) in terms of joint input-and-estimate distribution, the original estimation problem of MIMO nature was identical to that of a simple SISO estimation problem facing no self-interference (caused by the linear weighting), no nonlinear distortion, (caused by the random mapping), but only an effective AWGN;
2) the noise level of the above AWGN could be further determined by solving a set of coupled equations, whose dependency on the linear weighting and the random mapping was given explicitly;
3) as a byproduct of the replica analysis, the average MSE of the exact MMSE estimator could be computed directly from the fixed-point results (with no need for Mote Carlo simulations).
Comparing to existing works in the literature, this paper established a decoupling principle that not only extended the seminal work of \cite{GuoTIT2005RM} from 1L-SLM  to ML-GLM, but also indicated the universal existence of the principle in estimation under different models. As later shown in Part II, this decoupling principle carries great practicality and finds convenient uses in finite-size systems. To sum up, it opens a new avenue for the understanding and justification of the ML-GLM model, which is closely related to deep learning, or more precisely, to deep inference models such as the variational auto-encoder (VAE) \cite{fletcher2018inference}.

Replica method is not yet a rigorous method, and its justification is still an open problem in mathematical physics \cite{GuoTIT2005RM}. However, the method has evolved during the past $30$ years into a extremely powerful tool for attacking complicated theoretical problems as diverse as spin glasses, wireless communications, compressed sensing, protein folding, vortices in superconductors, and combinatorial optimization \cite{Mezard-Book09-InfoPhyCom}. Several of its important predictions have been confirmed by other rigorous approaches, e.g., the replica predictions for the SLM problem in \cite{GuoTIT2005RM} were verified in \cite{bayati2011dynamics} using a conditioning technique, and that for the GLM case \cite{schulke2016statistical} was very recently confirmed by \cite{Barbier-PNAS19-GAMP} through an interpolation approach. In this context, we referred to main results of this paper as claims and reminded the readers that their mathematical rigor are still pending on more breakthroughs.

Also, considering the implementation difficulty of an exact MMSE estimator, we continue to propose in Part II an approximate solution, whose computational complexity (per iteration) is as low as the GAMP, while its MSE performance is asymptotically Bayes-optimal.

\section{Acknowledgement}
This work was supported in part by the Scientific Research Fund of Guangzhou 201904010297, Natural Scientific Fund of Guangdong 2016A030313705, Special Fund for Applied Science and Technology of Guangdong 2015B010129001, and Natural Scientific Fund of Guangxi 2018GXNSFDA281013.

\ifCLASSOPTIONcaptionsoff
  \newpage
\fi


\end{document}